\documentclass[journal]{IEEEtran}
\usepackage[utf8]{inputenc}
\usepackage{bm}
\usepackage{epsfig,amsfonts,amsbsy,bm,mathrsfs}
\usepackage{amssymb,amsmath,amsthm,latexsym,amscd,amsfonts}
\usepackage{lettrine}
\usepackage{authblk}
\usepackage{graphics}
\usepackage{graphicx}
\usepackage{epstopdf}
\usepackage{psfrag,float}
\usepackage{pstricks}
\usepackage{pst-plot}
\usepackage{cite}
\usepackage{balance}
\usepackage{epsfig}
\usepackage{bbm}
\usepackage{enumitem}
\usepackage{dsfont}
\usepackage{setspace}
\usepackage{float}
\usepackage{relsize}
\usepackage{comment}
\usepackage{subfigure} 
\usepackage{algorithm}
\usepackage[noend]{algpseudocode}
\usepackage[nolist]{acronym}
\usepackage{tabularx}
\usepackage{pifont}
\usepackage{units}

\usepackage{tikz}
\usetikzlibrary{calc}
\usepackage{xcolor}
\usepackage{tabularx}
\usepackage{colortbl}
\usepackage{pgfplots}
\pgfplotsset{compat=newest}
\usetikzlibrary{plotmarks}
\usetikzlibrary{arrows.meta}
\usepgfplotslibrary{patchplots}
\usepackage{grffile}
\newlength\fheight 
\newlength\fwidth 
\usepgfplotslibrary{fillbetween}

\usepackage{amsthm}

\newtheorem{definition}{Definition}

\newtheorem{remark}{Remark}

\newcommand{\trp}{^\top}
\newcommand{\herm}{^\mathrm{H}}
\newcommand{\conj}{^{*}}

\algdef{SE}[SUBALG]{Indent}{EndIndent}{}{\algorithmicend\ }%
\algtext*{Indent}
\algtext*{EndIndent}

\makeatletter
\newcommand*\rel@kern[1]{\kern#1\dimexpr\macc@kerna}
\newcommand*\widebar[1]{%
  \begingroup
  \def\mathaccent##1##2{%
    \rel@kern{0.8}%
    \overline{\rel@kern{-0.8}\macc@nucleus\rel@kern{0.2}}%
    \rel@kern{-0.2}%
  }%
  \macc@depth\@ne
  \let\math@bgroup\@empty \let\math@egroup\macc@set@skewchar
  \mathsurround\z@ \frozen@everymath{\mathgroup\macc@group\relax}%
  \macc@set@skewchar\relax
  \let\mathaccentV\macc@nested@a
  \macc@nested@a\relax111{#1}%
  \endgroup
}
\makeatother

\newcommand{\rx}{{\text{R}}}

\newcommand{\Nrx}{N_\rx}

\newcommand{\Imatrix}{{ \boldsymbol{\mathrm{I}} }}
\newcommand{\boldzero}{{ {\boldsymbol{0}} }}
\newcommand{\yy}{\boldsymbol{y}}
\newcommand{\yylos}{\yy_{\rm{LOS}}}
\newcommand{\ww}{\boldsymbol{w}}
\newcommand{\YY}{\boldsymbol{Y}}

\newcommand{\DD}{\boldsymbol{D}}
\newcommand{\AAx}{\boldsymbol{A}_{{\rm{x}}}}
\newcommand{\Gammab}{\boldsymbol{\Gamma}}
\newcommand{\Phib}{\boldsymbol{\Phi}}
\newcommand{\Thetab}{\boldsymbol{\Theta}}
\newcommand{\Psib}{\boldsymbol{\Psi}}
\newcommand{\TT}{\boldsymbol{T}}

\newcommand{\XX}{\boldsymbol{X}}
\newcommand{\xx}{\boldsymbol{x}}
\newcommand{\MM}{\boldsymbol{M}}
\newcommand{\BB}{\boldsymbol{B}}
\newcommand{\UU}{\boldsymbol{U}}
\newcommand{\VV}{\boldsymbol{V}}
\newcommand{\JJ}{\boldsymbol{J}}
\newcommand{\Sigmab}{\boldsymbol{\Sigma}}
\newcommand{\RR}{\boldsymbol{R}}
\newcommand{\RRhat}{\widehat{\RR}}

\newcommand{\tauhat}{\widehat{\tau}}
\newcommand{\tauhatl}{\widehat{\tau}_{\ell}}
\newcommand{\tauhatlos}{\widehat{\tau}_{{\rm{LOS}}}}
\newcommand{\whatlos}{\widehat{\omega}_{{\rm{LOS}}}}
\newcommand{\whatmf}{\widehat{\omega}_{{\rm{MF}}}}
\newcommand{\tbhatlos}{\tbhat_{{\rm{LOS}}}}

\newcommand{\NN}{\boldsymbol{N}}
\newcommand{\yym}{\bm{\mathcal{Y}}}
\newcommand{\nnm}{\bm{\mathcal{N}}}
\newcommand{\dd}{\boldsymbol{d}}
\newcommand{\cc}{\boldsymbol{c}}

\newcommand{\nn}{\boldsymbol{n}}
\newcommand{\aab}{\boldsymbol{a}}
\newcommand{\aabx}{\aab_{\rm{x}}}
\newcommand{\aabz}{\aab_{\rm{z}}}
\newcommand{\es}{\sqrt{E_s}}
\newcommand{\deltaf}{\Delta_f}

\newcommand{\ddd}{\dd_S}

\newcommand{\nofdm}{N_{\text{OFDM}}}
\newcommand{\mband}{{{\rm{m}}}}
\newcommand{\ddm}{\dd_{\mband}}
\newcommand{\ddmtilde}{\widetilde{\dd}_{\mband}}
\newcommand{\ddmb}{\dd_{\mband,b}}
\newcommand{\ddmbg}[1]{\dd_{\mband,#1}}

\newcommand{\norm}[1]{\left\lVert#1\right\rVert}
\newcommand{\normsmall}[1]{\big\lVert#1\big\rVert}

\newcommand{\diag}[1]{ {\rm{diag}}\left(#1\right)  }
\newcommand{\rank}[1]{ {\rm{rank}}\left(#1\right)  }

\newcommand{\aabxw}{\widetilde{\aab}_{\rm{x}}}
\newcommand{\aabzw}{\widetilde{\aab}_{\rm{z}}}

\newcommand{\aabxhl}{\widehat{\aab}_{{\rm{x}},\ell}}
\newcommand{\aabzhl}{\widehat{\aab}_{{\rm{z}},\ell}}
\newcommand{\ddhl}{\widehat{\dd}_{\ell}}
\newcommand{\ddmhl}{\widehat{\dd}_{\mband,\ell}}
\newcommand{\ddmpq}{\ddm^{(p, q)}}
\newcommand{\ddmpqg}[2]{\ddm^{(#1, #2)}}

\newcommand{\yymt}{\widetilde{\yym}}

\newcommand{\YYm}{\widetilde{\YY}}
\newcommand{\NNm}{\widetilde{\NN}}

\newcommand{\UUm}{\widetilde{\UU}}
\newcommand{\Sigmabm}{\widetilde{\Sigmab}}
\newcommand{\VVm}{\widetilde{\VV}}

\newcommand{\aabxl}{{\aab}_{{\rm{x}},\ell}}
\newcommand{\aabzl}{{\aab}_{{\rm{z}},\ell}}
\newcommand{\ddl}{{\dd}_{\ell}}

\newcommand{\taul}{\tau_{\ell}}
\newcommand{\alphal}{\alpha_{\ell}}
\newcommand{\Mx}{N_{\rm{x}}}
\newcommand{\Mz}{N_{\rm{z}}}
\newcommand{\dz}{d_{\rm{z}}}
\newcommand{\dx}{d_{\rm{x}}}

\newcommand{\tbell}{\boldsymbol{t}_{\ell}}
\newcommand{\tb}{\boldsymbol{t}}
\newcommand{\tbhat}{\widehat{\tb}}

\newcommand{\hankel}{{\rm{Hankel}}}

\newcommand{\rmel}{{{\rm{el}}}}
\newcommand{\rmaz}{{{\rm{az}}}}
\newcommand{\rmlos}{{{\rm{LOS}}}}

\newcommand{\alphab}{\bm{\alpha}}
\newcommand{\taub}{\bm{\tau}}

\newcommand{\phib}{\bm{\phi}}
\newcommand{\phibl}{\phib_{\ell}}
\newcommand{\phibhat}{\widehat{\phib}}
\newcommand{\phiaz}{\phi_{\rmaz}}
\newcommand{\phiel}{\phi_{\rmel}}

\newcommand{\phihazl}{\widehat{\phi}_{\rmaz,\ell}}
\newcommand{\phihell}{\widehat{\phi}_{\rmel,\ell}}

\newcommand{\thetahazl}{\widehat{\theta}_{\rmaz,\ell}}
\newcommand{\thetahell}{\widehat{\theta}_{\rmel,\ell}}

\newcommand{\ellb}{\widehat{\ell}}
\newcommand{\tbhatel}{\tbhat_{\rmel}}

\newcommand{\thetahazlos}{\widehat{\theta}_{\rmaz,\rmlos}}
\newcommand{\thetahellos}{\widehat{\theta}_{\rmel,\rmlos}}

\newcommand{\thetaaz}{\theta_{\rmaz}}
\newcommand{\thetael}{\theta_{\rmel}}

\makeatletter
\newcommand{\gettikzxy}[3]{%
  \tikz@scan@one@point\pgfutil@firstofone#1\relax
  \edef#2{\the\pgf@x}%
  \edef#3{\the\pgf@y}%
}
\definecolor{mygray}{gray}{0.6}

\newcommand{\complexset}[2]{ \mathbb{C}^{#1 \times #2}  }
\newcommand{\realset}[2]{ \mathbb{R}^{#1 \times #2}  }
\newcommand{\complexsett}[3]{ \mathbb{C}^{#1 \times #2 \times #3}  }

\newcommand{\abs}[1]{\big\lvert #1 \big\rvert}

\begin{acronym}[ACRONYM]
\acro{AI}{artificial intelligence}
\acro{AoA}{angle-of-arrival}
\acro{AoD}{angle-of-departure}
\acro{AWGN}{additive white Gaussian noise}
\acro{BS}{base station}
\acro{BP}{belief propagation}
\acro{CDF}{cumulative density function}
\acro{CFO}{carrier frequency offset}
\acro{CP}{cyclic prefix}
\acro{CRB}{Cram\'er-Rao bound}
\acro{CCRB}{constrained Cram\'er-Rao bound}
\acro{C-V2X}{cellular vehicle-to-anything}
\acro{DA}{data association}
\acro{D-MIMO}{distributed multiple-input multiple-output}
\acro{DL}{downlink}
\acro{DSRC}{dedicated short-range communications}
\acro{EKF}{extended Kalman filter}
\acro{EM}{electromagnetic}
\acro{FIM}{Fisher information matrix}
\acro{GDOP}{geometric dilution of precision}
\acro{GNSS}{global navigation satellite system}
\acro{GPS}{global positioning system}
\acro{IP}{incidence point}
\acro{IQ}{in-phase and quadrature}
\acro{ISAC}{integrated sensing and communication}
\acro{ICI}{inter-carrier interference}
\acro{JCS}{Joint Communication and Sensing}
\acro{JRC}{joint radar and communication}
\acro{JRC2LS}{joint radar communication, computation, localization, and sensing}
\acro{IMU}{inertial measurement unit}
\acro{IOO}{indoor open office}
\acro{IoT}{Internet of Things}
\acro{IRN}{infrastructure reference node}
\acro{KPI}{key performance indicator}
\acro{LoS}{line-of-sight}
\acro{LS}{least-squares}
\acro{MCRB}{misspecified Cram\'er-Rao bound}
\acro{MIMO}{multiple-input multiple-output}
\acro{SIMO}{single-input multiple-output}
\acro{ML}{maximum likelihood}
\acro{mmWave}{millimeter-wave}
\acro{NLoS}{non-line-of-sight}
\acro{NR}{new radio}
\acro{OFDM}{orthogonal frequency-division multiplexing}
\acro{OTFS}{orthogonal time-frequency-space}
\acro{OEB}{orientation error bound}
\acro{PEB}{position error bound}
\acro{VEB}{velocity error bound}
\acro{PRS}{positioning reference signal}
\acro{QoS}{Quality of Service}
\acro{RAN}{radio access network}
\acro{RAT}{radio access technology}
\acro{RCS}{radar cross section}
\acro{RedCap}{reduced capacity}
\acro{RF}{radio frequency}
\acro{RIS}{reconfigurable intelligent surface}
\acro{RFS}{random finite set}
\acro{RMSE}{root mean squared error}
\acro{RSU}{road-side unit}
\acro{RTK}{real-time kinematic}
\acro{RTT}{round-trip-time}
\acro{SLAM}{simultaneous localization and mapping}
\acro{SLAT}{simultaneous localization and tracking}
\acro{SNR}{signal-to-noise ratio}
\acro{ToA}{time-of-arrival}
\acro{TDoA}{time-difference-of-arrival}
\acro{TR}{time-reversal}
\acro{TXRX}[TX/RX]{transmitter/receiver}
\acro{Tx}{transmitter}
\acro{Rx}{receiver}
\acro{UE}{user equipment}
\acro{UL}{uplink}
\acro{ULA}{uniform linear array}
\acro{URA}{uniform rectangular array}
\acro{UWB}{ultra wideband}
\acro{V2I}{vehicle-to-infrastructure}
\acro{V2X}{vehicle-to-anything}
\acro{V2V}{vehicle-to-vehicle}
\acro{VRU}{vulnerable road user}
\acro{CRU}{connected road user}
\acro{XL-MIMO}{extra-large MIMO}
\acro{REB}{range error bound}
\acro{MIMO}{multiple-input and multiple-output}
\acro{MPC}{multipath component}
\acro{WAA}{weighted average approximation}
\acro{SVD}{singular value decomposition}
\acro{ESPRIT}{estimation of signal parameters via rotational invariance techniques}
\acro{LTE}{long-term evolution}
\end{acronym}

\begin{document}

\bibliographystyle{IEEEtran}
\bstctlcite{IEEEexample:BSTcontrol}

\title{V2X Sidelink Positioning in FR1: Scenarios, Algorithms, and Performance Evaluation}
\author{Yu Ge,~\IEEEmembership{Student Member,~IEEE}, Maximilian Stark,~\IEEEmembership{Member,~IEEE}, Musa Furkan Keskin,~\IEEEmembership{Member,~IEEE}, Frank Hofmann, Thomas Hansen, Henk Wymeersch,~\IEEEmembership{Senior Member,~IEEE}\\
\thanks{Yu Ge, Musa Furkan Keskin, and Henk Wymeersch are with the Department of Electrical Engineering, Chalmers University of Technology, Sweden. (email: \{yuge,furkan,henkw\}@chalmers.se) \\
Maximilian Stark and Frank Hofmann are with Robert Bosch GmbH, Germany. (email:\{Maximilian.Stark2, Frank.Hofmann2). \\
This work was supported, in part,  by the Wallenberg AI, Autonomous Systems and Software Program (WASP) funded by Knut and Alice Wallenberg Foundation. 
}
}

\maketitle
\begin{abstract}
In this paper, we investigate sub-6 GHz V2X sidelink positioning scenarios in 5G vehicular networks through a comprehensive end-to-end methodology encompassing ray-tracing-based channel modeling, novel theoretical performance bounds, high-resolution channel parameter estimation, and geometric positioning using a \ac{RTT} protocol. We first derive a novel, approximate \ac{CRB} on the \ac{CRU} position, explicitly taking into account multipath interference, path merging, and the \ac{RTT} protocol. 
Capitalizing on tensor decomposition and ESPRIT methods, we propose high-resolution channel parameter estimation algorithms specifically tailored to dense multipath V2X sidelink environments, designed to detect \acp{MPC} and extract \ac{LoS} parameters. Finally, using realistic ray-tracing data and antenna patterns, comprehensive simulations are conducted to evaluate channel estimation and positioning performance, indicating that sub-meter accuracy can be achieved in sub-6 GHz V2X with the proposed algorithms.
\end{abstract}
\section{Introduction}
\acresetall 

The landscape of wireless communication is undergoing a profound transformation, with the  maturation of the fifth generation (5G) mobile radio systems and its successors, which promise to bring benefits for positioning and radar-like sensing \cite{wild2021joint},  exhibiting ever-growing importance in a diverse range of applications \cite{behravan-6g-2022}. Substantial attention has been dedicated to exploiting the millimeter-wave (mmWave) bands, 
due to their substantial bandwidth availability and corresponding superior distance resolution \cite{KanRap:21}. This empowers them to deliver superior data rates and high capacity, making them the go-to bands for many high-end applications, such as virtual reality and high-definition content streaming \cite{saad2019vision}. However, a promising segment lies in the lower bands, including the sub-6 GHz bands and 7-10 GHz spectrum. 
These bands hold great potential for precise positioning, especially in environments where multipath propagation is not overly complicated \cite{wild20236g}.  Applications include \ac{IoT}, public safety, and \ac{V2X} \cite{tr:38845-3gpp21}, where the latter is important to support a variety of driving functions that improve safety and comfort of passengers \cite{5g_nr_v2x_driving}. 


Within \ac{V2X}, sidelink communication can rely on the 5850-5925 MHz band at FR1, attributed to the  Intelligent Transportation Systems (ITS), and the neighboring unlicensed bands, to support  positioning  \cite{garcia2021tutorial,lien20203gpp}, augmenting its integrity, accuracy, and power efficiency \cite{tr:38859-3gpp22}. 
Due to its importance, the domain of sidelink positioning at sub-6 GHz  V2X has witnessed a surge of activities in terms of defining requirements, architectures, technologies, and possible standards for V2X high-accuracy positioning  \cite{liu2021highly,ganesan20235g,bartoletti2021positioning,shrivastava2023sidelink,decarli2023v2x}.
    \begin{figure}[tb]
        \centering 
        \subfigure[City scenario with vehicles and vulnerable road users.]{
    \includegraphics[width=0.4\textwidth]{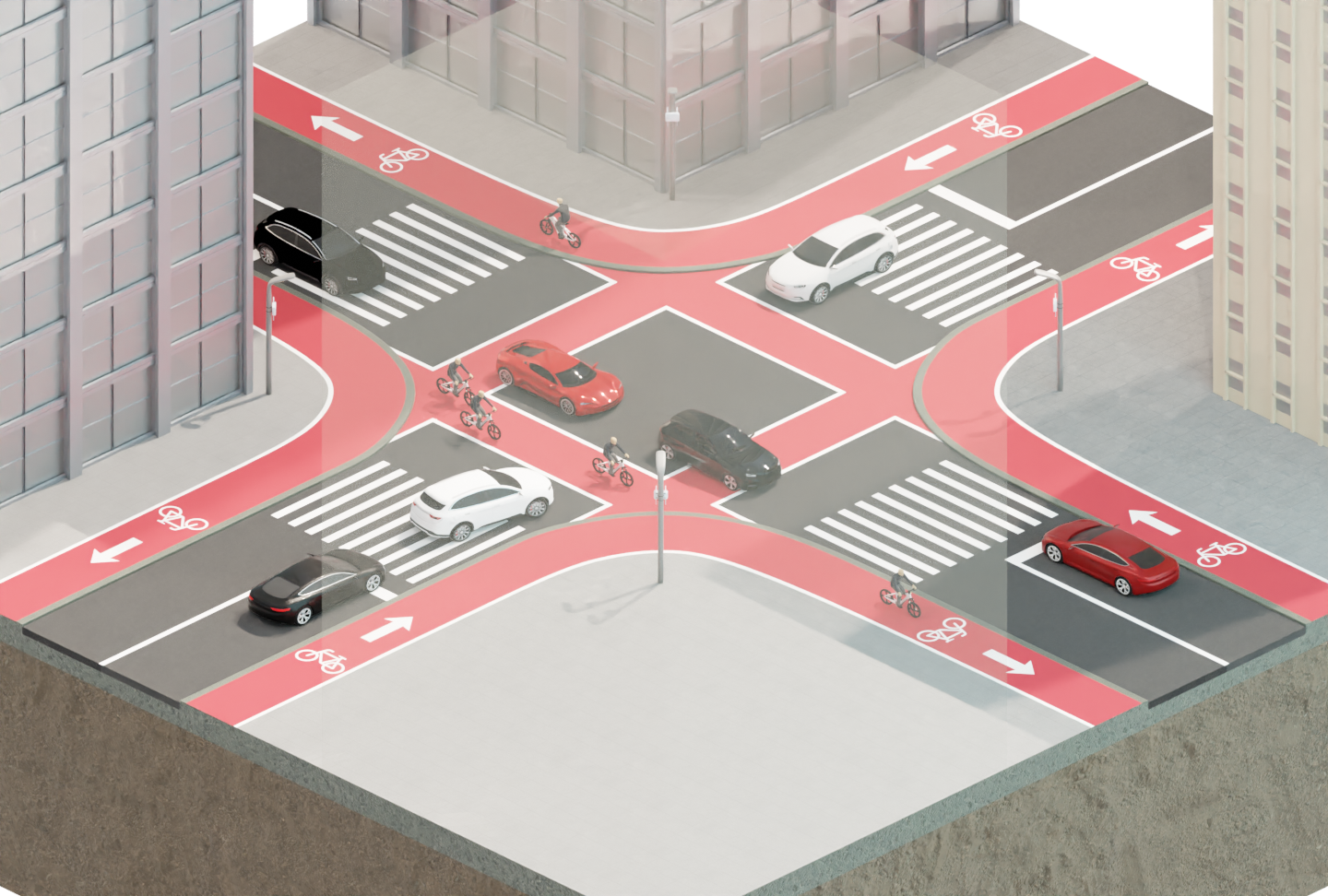} 
       \label{fig:scenarios-city}
        } 
        \\
        \subfigure[Highway scenario with only vehicles.]{
        \includegraphics[width=0.4\textwidth]{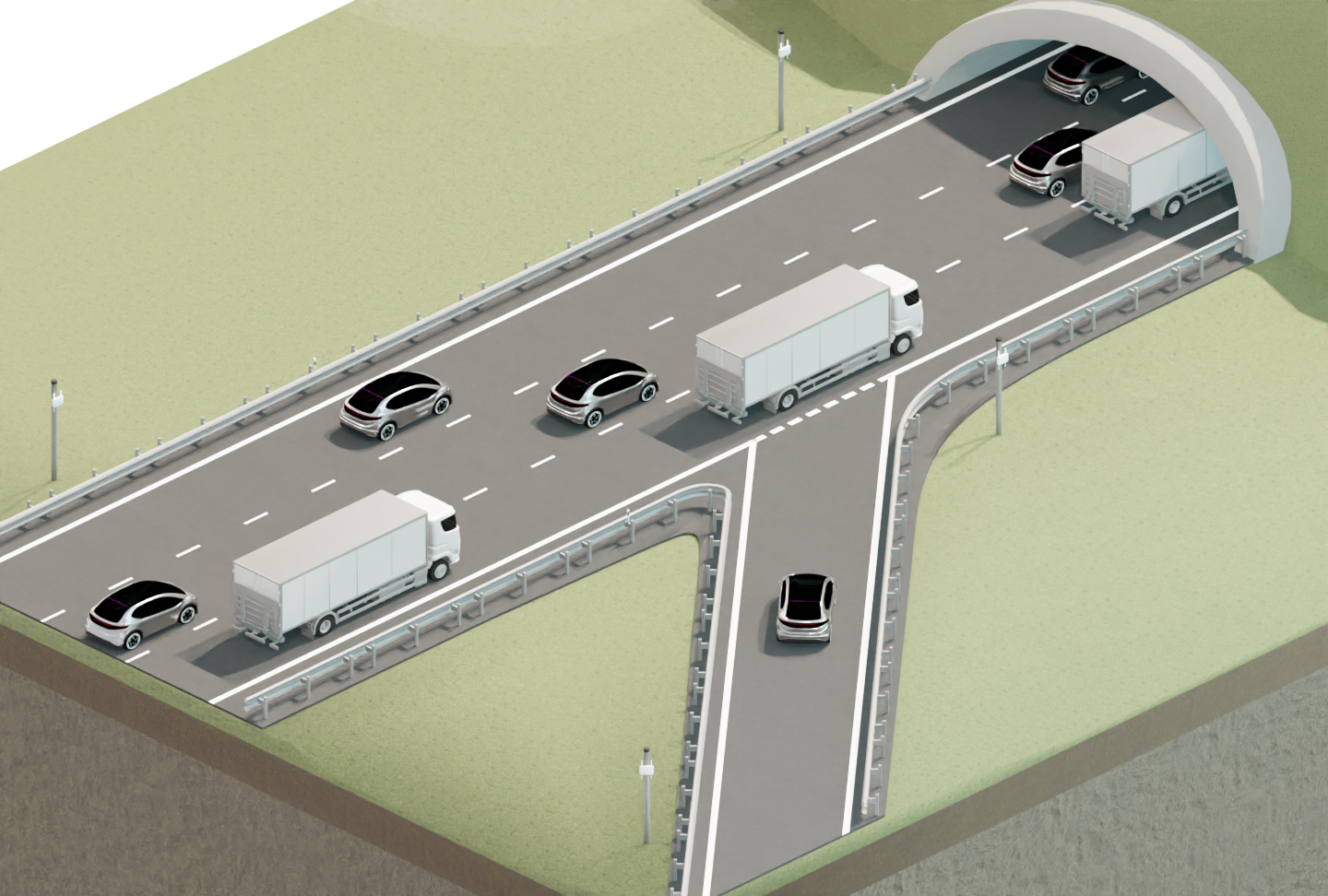}
        \label{fig:scenarios-hwy}
        }
        \caption{Two scenarios considered in this work for \ac{V2X} sidelink positioning at FR1. The city scenario features a more complex propagation environment, while the highway scenario is subject to blockage from trucks and higher speeds. }
        \label{fig:scenarios}
    \vspace{-5mm} \end{figure}
Technical research has  covered different aspects of the sidelink \ac{V2X} positioning problem, and two exemplifying scenarios are shown in Fig.~\ref{fig:scenarios}. 
Papers can be divided into two categories: those that start from processed range or angle measurements \cite{ma2019efficient,liu2021v2x,fouda2021dynamic,hossain2017high} and those that also consider how to obtain such measurements \cite{del2017performance,kakkavas2018multi,JCS_V2X_2022,ge2023analysis,chen2023multi,ammous2023zero}. 
In the former category, advances are mainly in the post-processing and fusion of different measurements: 
\cite{ma2019efficient} proposes a method  for positioning from a single \ac{RSU} using range measurements over time; 
\cite{liu2021v2x} combines \ac{V2V} range and angle, and \ac{V2I} \ac{TDoA} measurements are combined to improve positioning;  in \cite{fouda2021dynamic}, a dynamic method is proposed to switch between \ac{GNSS} and  NR \ac{V2X} \ac{TDoA} measurements; \cite{hossain2017high} solely relies  on \ac{AoA} measurements, derived from ray-tracing data. 
In the latter category, we find studies on the statistical evaluation of the measurement errors for \ac{LTE} \cite{del2017performance} and 5G \cite{JCS_V2X_2022,ge2023analysis}; 
\ac{CRB} analyses of positioning based on \ac{AoA} and \ac{TDoA} measurements at FR1 and FR2; and more futuristic scenarios in  \cite{chen2023multi,ammous2023zero} that combine sidelink with one or several \acp{RIS}, in order to perform absolute positioning or several \acp{UE} without \acp{BS}  or \acp{RSU}.

Surprisingly, few studies have addressed the problem of estimating channel parameters (i.e., delays, angles, Dopplers) needed for positioning in the V2X sidelink positioning context, where antenna arrays are small, bandwidth is limited, and possibly spread over several sub-bands, and multipath is rich and performance-limiting. Generic channel parameter estimation methods can be grouped into 3 categories: (i) methods based on \ac{ML}, such as SAGE \cite{fleury1999channel}, which have excellent performance but high complexity; (ii) sparsity-based methods, such as compressive sensing \cite{bajwa2010compressed}, which have high complexity in  large dimensions; and (iii) subspace-based methods such as \ac{ESPRIT} \cite{viberg1991sensor}, which avoid high-dimensional searches and can provide a favorable performance-complexity trade-off. Closely-related papers that cover the salient properties of sidelink \ac{V2X} (small arrays, limited bandwidth spread over several sub-bands, and rich multipath) include \cite{tensor_MIMO_OFDM_2022} for single-band processing and 
\cite{delayTarik_TWC_2022,multiband_delay_TWC_2023,multiband_superresolution_ahmad_2022} for multi-band processing. In particular, \cite{tensor_MIMO_OFDM_2022} considers MIMO-OFDM and develops a multipath channel estimation method to estimate \ac{AoA}, \ac{AoD} and \ac{ToA} of multiple paths from 3-D tensor observations. To overcome the small number of antenna elements, a spatial augmentation method is proposed to move resources from the frequency domain to the spatial domains. This enables accurate detection of the number of paths and estimation of  path parameters in rich multipath environments. The algorithm has been verified using experimental data. In terms of multi-band processing, all prior works have focused exclusively on delay estimation. In particular, 
\cite{delayTarik_TWC_2022} proposes and experimentally validates a subspace-fitting approach at sub-6 GHz; the solution from \cite{multiband_superresolution_ahmad_2022} is based on the MUSIC algorithm, and  \cite{multiband_delay_TWC_2023} proposes  a two-stage delay estimation scheme, first ignoring and then utilizing phase coherence across multiple bands.

It is clear that despite  expansive research activities, several gaps remain. First, there is an urgent need to understand the performance of sub-6 GHz positioning for vehicular use cases in realistic environments with challenging multipath propagation, considering both angle and delay measurements, as well as the salient sidelink V2X properties. Second, there is a lack of methods that can perform channel parameter estimation and operate well under the harsh conditions faced by \ac{V2X} sidelink. Finally, the third gap pertains to the impact of different antenna and bandwidth configurations. Depending on these configurations, the resolution and accuracy of positioning can be significantly affected. Understanding these relationships and how to optimize these configurations for improved multipath robustness and enhanced positioning accuracy is a crucial area that needs more exploration.

In this work, we address these three gaps pertaining to the applicability and performance of sidelink positioning in FR1 for automotive applications. Through an end-to-end approach that spans ray-tracing-based channels, high-resolution channel parameter estimation, and geometric positioning using a \ac{RTT} protocol, we aim to elucidate the circumstances under which sidelink positioning can fulfill automotive use case requirements. Our methodology explores diverse operating conditions, analyzing the impact of variables such as antenna patterns, the number of antenna elements, multiband processing, and the influence of prior information. This work expands on our previous conference contribution \cite{ge2023analysis} and presents the following significant contributions:
\begin{itemize}
    \item \textbf{Performance bounds for sidelink positioning under multipath:} 
    A novel, approximate \ac{CRB} on the \ac{UE} position is derived, taking into account multipath interference and the \ac{RTT} protocol. The proposed \ac{CRB} builds upon and generalizes the bound presented in \cite{ge2023analysis}, revealing multipath resolution as a primary limiting factor. This finding underscores the necessity for the development of super-resolution methods specifically tailored to this context.

    

    \item \textbf{High-resolution multi-band channel parameter estimation algorithms for V2X sidelink:} To address the distinctive challenges of V2X sidelink scenarios, which involve low bandwidth, small arrays and dense multipath, we propose novel high-resolution, multi-dimensional and multi-band channel estimation algorithms based on tensor decomposition and ESPRIT to detect \acp{MPC}, estimate their delay, azimuth and elevation parameters, and
extract the \ac{LoS} path. The proposed algorithms leverage the multi-band shift invariance structure of large frequency apertures \cite{multESP}, as well as a spatial augmentation strategy \cite{tensor_MIMO_OFDM_2022} specifically tailored to multi-band operation, to resolve \acp{MPC} for both \acp{URA} and \acp{ULA} under the challenging conditions of V2X settings. 

    \item \textbf{End-to-end holistic performance evaluation and validation:}
    We conduct a comprehensive numerical study employing realistic channel impulse responses from ray tracing, antenna patterns, and transmission powers to gauge the achievable performance of the proposed processing chain under various scenarios. Additionally, we assess the utility of different types of prior information (e.g., orientation information) in reducing location uncertainty. These analyses enable us to determine the conditions under which sidelink positioning can support key vehicular use cases.
\end{itemize}

\section{System Model} \label{sec:systemmodel}
In this section, we introduce the state models of the \ac{RSU}, \ac{CRU}, and \acp{IP} in the considered sub-6 GHz V2X sidelink scenarios. In addition, the received signal model and the measurement model are also presented.
%
We consider sub-6 GHz V2X sidelink scenarios with  one \ac{RSU} and one \ac{CRU}. These devices are not synchronized with each other. The \ac{RSU} is equipped with a \ac{URA} or a \ac{ULA}, with its center located at $\boldsymbol{x}_{\text{RSU}}=[x_\text{RSU},y_\text{RSU},z_\text{RSU}]^{\textsf{T}}$. In addition, it has a rotation $\boldsymbol{\psi}_\text{RSU}=[\varepsilon_\text{RSU},\varpi_\text{RSU},\gamma_\text{RSU}]^{\textsf{T}}$,  denoting the roll, pitch, yaw with respect to the global reference coordinate system, respectively. Therefore, we can describe the state of the \ac{RSU} as $\boldsymbol{s}_\text{RSU}=[\boldsymbol{x}_\text{RSU}^{\mathsf{T}},\boldsymbol{\psi}_\text{RSU}^{\mathsf{T}}]^{\mathsf{T}}$, which is known and fixed over time. 
The 
\acp{CRU} is located at $\boldsymbol{x}_{\text{CRU}}=[x_{\text{CRU}},y_{\text{CRU}},z_{\text{CRU}}]^{\textsf{T}}$. 

\subsection{Received Signal Model} \label{received signal model}
Even though a transmitter may have multiple antennas, only a single transmit antenna (or a single beam) is used for transmission. This allows us to work with a \ac{SIMO} case with   $\Nrx$ antenna elements at the receiver sides, which is suitable to capture both links in an \ac{RTT} protocol (described in Section \ref{sec:RTTProtocol}).  
At each time step, the \ac{RSU} sends signals to the \ac{CRU}. Those signals can reach the \ac{CRU} via \ac{LoS} path, or \ac{NLoS} paths, or both. The \ac{LoS} path represents the signals reaching the receiver directly, while the \ac{NLoS} paths represent the signals reaching the receiver indirectly, reflected or scattered by objects in the environment. The \ac{CRU} responds with a signal to the \ac{RSU}, according to the \ac{RTT} protocol. 
We consider multiband \ac{OFDM} transmission over $B$ separate frequency bands, where $f_b$ is the carrier frequency of the $b$-th band, and $N_{\text{OFDM}}$ and $S$ denote the number of symbols and subcarriers in each band, respectively. 
We will assume that the transmission interval is sufficiently short, so that the Doppler-induced phase can be  ignored. 
This  yields a model, after coherent integration over $\nofdm$ symbols and wiping off the unit-modulus pilots, of the form \cite{heath2016overview}
\begin{align} 
     &\boldsymbol{y}_{s,b}= \sqrt{E_s} \sum_{{\ell}=0}^{L-1}\alphal \boldsymbol{a}(\boldsymbol{\theta}_{{\ell}})
     e^{-\jmath 2\pi (f_b + s \Delta_f) \taul}    + \boldsymbol{n}_{s,b} ~,\label{eq:reveivedsignalCompressed}
\end{align}
where $E_s = P N_{\text{OFDM}}/W$ denotes the energy per subcarrier after coherent integration, in which $P$ is the average transmit power and $W = S B \deltaf$ denotes the total bandwidth. In addition, 
 $\boldsymbol{n}_{s,b} \sim \mathcal{CN}(\bm{0},N_0\bm{I})$ is the \ac{AWGN},  
 $\Delta_f$ is the subcarrier spacing, $T$ is the OFDM symbol duration (including \ac{CP}), and $\lambda$ is the wavelength. We emphasize that the model \eqref{eq:reveivedsignalCompressed} is valid at both the \ac{RSU} and \ac{CRU} side. In \eqref{eq:reveivedsignalCompressed}, there are $L$ paths, where each path is either the \ac{LoS} path  (index $\ell=0$) or a \ac{NLoS} path (index $\ell>0$). 
Each path can be characterized by a complex gain $\alphal$, a \ac{ToA} $\tau_{\ell}$, 
and an \ac{AoA} $\boldsymbol{\theta}_{{\ell}}$.  The \ac{AoA} has azimuth and elevation components, i.e., $\boldsymbol{\theta}_{{\ell}}=[\theta_{\text{az},{\ell}},\theta_{\text{el},{\ell}}]^{\mathsf{T}}$. 

\subsection{Geometric Relations}

The \ac{AoA} $\boldsymbol{\theta}_{\ell}=[\theta_{\text{az},\ell},\theta_{\text{el},\ell}]^{\textsf{T}}$ is determined by the arrival direction of the signals at the receiver side (\ac{CRU} or \ac{RSU}), denoted as $\boldsymbol{t}_{\ell}$, which is considered in the local coordinate system of the receiver and can be calculated by 
\begin{align}
   \boldsymbol{t}_{
   \ell} = \boldsymbol{R}_\rx\frac{\boldsymbol{x}_{\text{IP},\ell}^{*}-\boldsymbol{x}_\rx}{\Vert\boldsymbol{x}_{\text{IP},\ell}^{*}-\boldsymbol{x}_\rx \Vert}\label{eq:directional_vector}
\end{align}
where $\boldsymbol{R}_\rx$ is the rotation matrix from the global reference coordinate system to the local coordinate system of the receiver, $\boldsymbol{x}_{\text{IP},\ell}^{*}$ is the 3D position of the last \ac{IP} of the given path, which is the point where the signal lastly hits the landmark (it is the position of the receiver $\boldsymbol{x}_\rx$ if the path is the \ac{LoS} path). The \ac{AoA} $\boldsymbol{\theta}_{\ell}$ is given by $ \theta_{\text{az},\ell} = \mathrm{arctan2}([\boldsymbol{t}_{\ell}]_{1},[\boldsymbol{t}_{\ell}]_{2})$ and $ \theta_{\text{el},\ell} = \arcsin([\boldsymbol{t}_{\ell}]_{3},\Vert\boldsymbol{t}_{\ell}\Vert)$. 
It is clear that we can parameterize the \ac{AoA} by the unit vector $\Vert \boldsymbol{t}_{\ell} \Vert =1$ via
\begin{align}
 \boldsymbol{t}_{\ell}=[\cos (\theta_{\text{el},\ell})\sin (\theta_{\text{az},\ell}) ~\cos (\theta_{\text{el},\ell})\cos (\theta_{\text{az},\ell}) ~ \sin(\theta_{\text{el},\ell})]^\top  \label{eq_ttb}
\end{align}
and both notations will be used interchangeably. 
The \ac{ToA} $\tau_{\ell}$ in each link of the RTT protocol is given by $ \tau_{\ell} = {d_{\ell}}/{c}+\beta$, 
where $d_{\ell}$ is the propagation distance, which for  the \ac{LoS} path ($\ell=0$) is $d_{0}=\Vert\boldsymbol{x}_\text{CRU}-\boldsymbol{x}_\text{RSU}\Vert$, $c$ is the speed of light, and $\beta$ is the clock bias between the \ac{RSU} and \acp{CRU}. 

\subsection{RTT Protocol and Positioning Inputs} \label{sec:RTTProtocol}

The signal model \eqref{eq:reveivedsignalCompressed} is valid for both the \ac{RSU} and \ac{CRU} sides when acting as a receiver. However, \ac{ToA} obtained from \eqref{eq:reveivedsignalCompressed} can be converted to distance only under perfect synchronization between the transmitter and the receiver. In the absence of synchronization, delay measurements involve the combined impact of distance and relative clock offset $\beta$. 
RTT offers a solution to deal with the clock offset in range-based positioning \cite{5g_nr_rel16,5g_nr_v2x_driving}. 
The \ac{RSU} can measure the round-trip time to the \ac{CRU} and infer the distance by subtracting the known processing time at the \ac{CRU}.
In particular, the  \ac{ToA} estimate at the requester (the \ac{RSU}) is given by $ \hat{\tau}_{0,\text{RSU}} = t_{\text{res}} + \Vert \bm{x}_{\text{CRU}}-\bm{x}_{\text{RSU}} \Vert/c +n_{\text{RSU}}$, 
where $t_{\text{res}}$ is the time at which the response was sent (in the clock of the requester/\ac{RSU}) and $n_{\text{RSU}}$ is measurement noise.
The response time $t_{\text{res}}$ depends on the \ac{ToA} estimate at the responder/\ac{CRU} and is given by  $t_{\text{res}}  =\hat{\tau}_{0,\text{CRU}}-\beta$, in which the \ac{ToA} estimate is 
$\hat{\tau}_{0,\text{CRU}}=t_{\text{req}}+ \Vert \bm{x}_{\text{CRU}}-\bm{x}_{\text{RSU}} \Vert/c + \beta +n_{\text{CRU}}$, 
where $t_{\text{req}}$ is the time at which the request was sent and $\beta$ is the clock bias of the responder and $n_{\text{CRU}}$ is the ToA measurement noise at the \ac{CRU}. When combined, we find that 
\begin{align}
      \hat{\tau}_{0,\text{RSU}} =t_{\text{req}}+ 2\Vert \bm{x}_{\text{CRU}}-\bm{x}_{\text{RSU}} \Vert/c + n_{\text{RSU}}+n_{\text{CRU}}, \label{eq:RTTobservation}
\end{align}
from which the distance is then easily recovered.

As we only consider the channel parameters of the \ac{LoS} paths for positioning, the final observation is 
 $\boldsymbol{z}_{0}=[\hat{\tau}_{0,\text{RSU}},\hat{\boldsymbol{\theta}}^{\mathsf{T}}_{0,\text{RSU}}]^{\mathsf{T}}$. By summarizing \eqref{eq:directional_vector}--\eqref{eq:RTTobservation}, we have the measurement function of the \ac{LoS} path as  $   \boldsymbol{z}_{0}= \boldsymbol{h}(\boldsymbol{x}_\text{CRU},\boldsymbol{x}_\text{RSU})+\boldsymbol{r}_{0}$, 
where $\boldsymbol{h}(\boldsymbol{x}_\text{CRU},\boldsymbol{x}_\text{RSU})$ is the nonlinear function that performs geometric mapping from $\boldsymbol{x}_\text{CRU}$, $\boldsymbol{x}_\text{RSU}$ to $\boldsymbol{z}_{0}$, and  $\boldsymbol{r}_{0}\sim \mathcal{N}\left(\boldsymbol{0},\boldsymbol{R}_{0}\right)$ is the measurement noise with $\boldsymbol{R}_{0}$ denoting the covariance. For a fair comparison of different deployments, we assume that the height of \ac{CRU} (i.e., $z_\text{CRU}$) is known, as the \ac{CRU} always moves on the ground in the considered scenarios. 
Then, only X and Y-coordinate of $\boldsymbol{x}_\text{CRU}$, $x_\text{CRU}$ and $y_\text{CRU}$, are unknown, and need to be estimated. Given the measurement  $\boldsymbol{z}_{0}$ for the \ac{LoS} path, and covariance $\boldsymbol{R}_{0}$, the \ac{ML} estimate of the unknown $x_\text{CRU}$ and $y_\text{CRU}$ can be obtained via 
\begin{align} \label{eq_ml}
    &\widehat{x}_\text{CRU},\widehat{y}_\text{CRU} = \\ &
    \arg \min_{x_\text{CRU},y_\text{CRU}} \,  (\boldsymbol{z}_{0} - \boldsymbol{h}(\boldsymbol{x}_\text{CRU},\boldsymbol{x}_\text{RSU}))^{\mathsf{T}} \boldsymbol{R}_{0}^{-1} (\boldsymbol{z}_{0} - \boldsymbol{h}(\boldsymbol{x}_\text{CRU},\boldsymbol{x}_\text{RSU})),\nonumber
\end{align}
which can be solved via gradient descent, and its initial point can be obtained via \ac{LS} approaches in 
closed-form \cite{zhu2009simple}.





\section{Fundamental Performance Bounds}

 In this section, we extend the bound from \cite{ge2023analysis} by considering angle measurements as well as the \ac{RTT} protocol, to determine an approximate \ac{CRB} on the position of the responder in the frame of the reference of the initiator. Since the number of paths is not known and paths may not be mutually resolvable, the \ac{CRB} 
requires careful consideration.  
\subsection{Multi-dimensional Resolution Cell and Path Merging}
To account for the finite resolution of channel parameter estimation methods, we define a multi-dimensional resolution cell, based on the standard concepts of range and angle resolution. 
\begin{definition}[Multi-dimensional resolution cell]
Consider a receiver with $N_{\text{R}}$ elements with positions $\boldsymbol{p}_{m}=[p_{m,x},0,p_{m,z}]^{\top}$ in the local coordinate frame and an OFDM system with $S$ subcarriers and subcarrier spacing $\Delta_f$. 
    Given a reference path with delay $\tau$ and arrival direction $\bm{t}=[t_x,t_y,t_z]^\top$, 
    the resolution cell $\mathcal{R}(\tau,\bm{t})$ with respect to the reference path is the set of all channel parameters $\tilde{\tau},\tilde{\bm{t}}$ such that (i) $|\tau-\tilde{\tau}|<1/(S\Delta_{f})$; (ii) $|t_{x}-\tilde{t}_{x}|<\lambda / |\max_{m}p_{m,x}-\min_{m}p_{m,x}|$; (iii) $|t_{z}-\tilde{t}_{z}|<\lambda / |\max_{m}p_{m,z}-\min_{m}p_{m,z}|$. In other words, $(\tilde{\tau},\tilde{\bm{t}})\in \mathcal{R}(\tau,\bm{t})$ when the three conditions (i), (ii), (iii) hold simultaneously.  \label{def:resolutionCell}
\end{definition}
From Def.~\ref{def:resolutionCell}, the \ac{LoS} resolution cell is then given by $\mathcal{R}(\tau_0,\bm{t}_0)$. We introduce a reduced set of paths $\mathcal{L}$, comprising only the paths within the \ac{LoS} resolution cell, i.e., $\ell\in \mathcal{L} \iff (\tau_{\ell},\bm{t}_{\ell})\in \mathcal{R}(\tau_0,\bm{t}_0)$. From this, we can define an observation model  with a reduced number of paths, removing all the non-informative paths that are resolvable with respect to the \ac{LoS} from \eqref{eq:reveivedsignalCompressed}, i.e., summing only over $\ell\in \mathcal{L}$.
Finally, we introduce the concept of merged paths, which corresponds to how the observation would appear to non-super-resolution methods. 
\begin{definition}[Path merging] \label{def:merging}
Given the observation model \eqref{eq:reveivedsignalCompressed} and the set of paths $\mathcal{L}$ that fall within the \ac{LoS} resolution cell $\mathcal{R}(\tau_0,\bm{t}_0)$, the observation as seen by non-super-resolution methods can be considered as a single merged path
$ \boldsymbol{y}_{s,b}= \sqrt{E_s} \bar{\alpha}_{0} \boldsymbol{a}(\bar{\boldsymbol{\theta}}_{0})
     e^{-\jmath 2\pi (f_b + s \Delta_f) \bar{\tau}_{0}}    + \boldsymbol{n}_{s,b} $,
 where $  \bar{\alpha}_0 = \sum_{\ell \in \mathcal{L}} \alpha_{\ell}$,  $\bar{\tau}_0 = \sum_{\ell \in \mathcal{L}} w_{\ell}\tau_{\ell} $,  and $\bar{\boldsymbol{\theta}}_{0}=\sum_{{\ell} \in \mathcal{L}} w_{\ell}\boldsymbol{\theta}_{\ell} $. The weights $w_{\ell}$ are computed based on the relative path amplitudes, i.e.,  $w_{\ell}={|\alpha_{\ell}|}/{\sum_{{\ell}' \in \mathcal{L}}|\alpha_{{\ell}'}|}$.

\end{definition}

\subsection{\ac{CRB} of Channel Parameters}
On either the request or the response transmission in the RTT protocol, the signal model is of the same form, so the \ac{CRB} of the channel parameters can be computed in a unified way. 
We introduce the parameters of the merged path as
$\bar{\bm{\xi}}=[\bar{\boldsymbol{t}}_{0}^{\top},\bar{\tau}_{0},\Re\bar{\alpha}_{0},\Im\bar{\alpha}_{0},]^{\top}\in\mathbb{R}^{6}$, which will be estimated by non-super-resolution methods,\footnote{Note that the case where the \ac{LoS} path is resolved from the \ac{NLoS} paths is a special case with $w_0=1$. Hence, the path merging procedure does not affect the observation when the \ac{LoS} is resolvable from the other paths. } with bias $\boldsymbol{b}=\bar{\boldsymbol{\xi}}-\boldsymbol{\xi}$, where ${\bm{\xi}}=[{\boldsymbol{t}}_{0}^{\top},{\tau}_{0},\Re{\alpha}_{0},\Im{\alpha}_{0},]^{\top}\in\mathbb{R}^{6}$ denotes the \ac{LoS} path parameters.
We compute the $6\times6$ FIM  $\boldsymbol{J}(\bar{\bm{\xi}})$ 
from the model  in Def.~\ref{def:merging}~\cite{kay1993fundamentals}. 
We subsequently remove any unobserved dimensions  to obtain a full-rank FIM.\footnote{In particular, we always remove $\bar{t}_{y,0}$ since arrays always lie on the local XZ-plane, we additionally remove $\bar{t}_{z,0}$ when the receiving array is a linear array, and we additionally remove $\bar{t}_{x,0}$ when the receiver has no array. The size of $\bar{\bm{\xi}}$ is adjusted correspondingly, as are the sizes of $\bm{\xi}$ and the bias $\bm{b}$. }
Hence, the lower bound on the error covariance of 
${\bm{\xi}}$ is given by 
\begin{align}
    \mathbb{E}\{(\hat{\bm{\bar{\xi}}}-\boldsymbol{\xi})(\hat{\bm{\bar{\xi}}}-\boldsymbol{\xi})^{\top}\}\ge\boldsymbol{J}^{-1}(\bar{\boldsymbol{\xi}})+\boldsymbol{b}\boldsymbol{b}^{\top}. \label{eq:biasedCRB}
\end{align}

\subsection{CRB of the Position Parameters} \label{sec:CRB-proposed}
From the request transmission, we obtain the bound from \eqref{eq:biasedCRB} on the \ac{ToA} $\hat{\tau}_{0,\text{CRU}}$ as $\sigma^2_{\text{ToA}}$. From the response transmission, we also obtain the bound from \eqref{eq:biasedCRB} 
on the \ac{ToA} $\hat{\tau}_{0,\text{RSU}}$ and the observable components of the arriving direction $\boldsymbol{t}_{0,\text{RSU}}$, as $\bm{\Sigma} \in \mathbb{R}^d$, where $d=2$ for a \ac{RSU} with a linear array and $d=3$ for a \ac{RSU} with a planar array. Recalling \eqref{eq:RTTobservation}, 
the channel parameters $\bm{\zeta}= [{\tau}_{0,\text{RTT}},[\bm{t}_{0,\text{RSU}}]^\top_{1:d}]^\top$ will have covariance matrix of the observable components 
\begin{align}
    \tilde{\bm{\Sigma}}=\bm{\Sigma} + \sigma^2_{\text{ToA}} \bm{e}_1 \bm{e}^\top_1 \in \mathbb{R}^{d \times d}, \label{eq:WAA}
\end{align}
where $\bm{e}_1=[1,\bm{0}^\top_{d-1}]^\top$ and ${\tau}_{0,\text{RTT}}=2\Vert \bm{x}_{\text{CRU}}-\bm{x}_{\text{RSU}} \Vert/c$. Using a standard Jacobian approach, the position error bound on the \ac{CRU} location can be computed. Note that when $d=2$, the height of the \ac{CRU} must be known to localize the \ac{CRU}. 
In summary, we  can compute three different bounds:
\begin{itemize}
    \item \emph{The \ac{LoS} \ac{PEB}}, where in \eqref{eq:reveivedsignalCompressed} only the path $\ell=0$ is maintained, in both links of the \ac{RTT} protocol.  
    \item \emph{The \ac{NLoS} \ac{PEB}}, where in \eqref{eq:reveivedsignalCompressed} all the paths $\ell\in \mathcal{L}$ in the \ac{LoS} resolution cell are maintained, in both links of the \ac{RTT} protocol. 
    \item \emph{The \ac{WAA}}, where in \eqref{eq:reveivedsignalCompressed} the paths $\ell\in \mathcal{L}$ are merged and the bound is found via 
    \eqref{eq:WAA}. 
\end{itemize}

\section{Algorithms for Channel Parameter Estimation and Positioning} \label{Sec:ChannelEstimation}
In this section, we tackle the problem of estimating the \ac{LoS} parameters, specifically the delay $\tau_0$ and the azimuth and elevation \acp{AoA} $\theta_{\rmaz,0}$ and $\theta_{\rmel,0}$, using the observation \eqref{eq:reveivedsignalCompressed}. 
Multipath channel estimation in sub-6 GHz V2X sidelink scenarios presents distinctive challenges due to factors such as small arrays (consisting of few antenna elements) \cite{tr:37885-3gpp19}, narrow bandwidth (around $10$ MHz) \cite{ko2021v2x} and dense multipath conditions \cite{ganesan20235g}. To overcome these challenges, we propose channel parameter estimation algorithms specifically tailored to V2X sidelink environments, designed to detect \acp{MPC} in \eqref{eq:reveivedsignalCompressed} and extract \ac{LoS} parameters. Our proposed algorithms encompass three different array configurations: \ac{URA}, \ac{ULA}, and single-antenna setups, leading to variations in the structure of the array steering vector $\aab(\cdot)$ in \eqref{eq:reveivedsignalCompressed}. 





\subsection{URA: 3-D Tensor Observations}\label{sec_3d_obs}
When the receiver is equipped with a \ac{URA}, the observation in \eqref{eq:reveivedsignalCompressed} can be reshaped into a 3-D tensor as
\begin{align} \label{eq_yy_3d}
     \yym = \es \sum_{\ell=0}^{L-1}\alphal \aabz(\tbell) \circ \aabx(\tbell) \circ \dd(\taul) + \nnm \in \complexsett{\Mz}{\Mx}{SB}, 
 \end{align}
 where $\circ$ denotes the outer product. In \eqref{eq_yy_3d}, the horizontal and vertical spatial-domain steering vectors, $\aabx(\tb) \in \complexset{\Mx}{1}$ and $\aabz(\tb) \in \complexset{\Mz}{1}$, are defined, respectively, as
 \begin{align} \label{eq_ax}
    [\aabx(\tb)]_n &= e^{\jmath \frac{2 \pi}{\lambda}   \dx n \cos(\thetael) \sin(\thetaaz)} ~,
    [\aabz(\tb)]_n 
    = e^{\jmath \frac{2 \pi}{\lambda}   \dz n \sin(\thetael)} ~,
\end{align}
and  $\tb_\ell$ was introduced in \eqref{eq_ttb}. 
Here, $\dx$ and $\dz$ denote the element spacing, and $\Mx$ and $\Mz$ are the number of antenna elements in the horizontal and vertical axes of the \ac{URA} respectively, with $\Mx \Mz = \Nrx$ and $\aab(\tb) = \aabx(\tb) \otimes \aabz(\tb)$.
The goal is to estimate the parameters of multiple paths from \eqref{eq_yy_3d} and resolve the \ac{LoS} component from the \ac{NLoS} paths under dense multipath conditions typically arising in \ac{V2X} sidelink scenarios \cite{ganesan20235g}. To this end, we propose three multi-band channel estimation algorithms, with the first two serving as benchmarks and the last one designed specifically to address the unique challenges of \ac{V2X} sidelink settings. 

\subsubsection{Matched Filtering (MF) Benchmark}
We devise a low-complexity matched filtering (MF) approach \cite{gezici_UWB_2005} that first estimates the \ac{LoS} delay via non-coherent integration across the \ac{URA} elements and then retrieves the \ac{LoS} angles by using this delay estimate. Specifically, after recasting \eqref{eq_yy_3d} into a matrix form, i.e.,
\begin{align} \label{eq_yy_2d}
     \YY = \es \sum_{\ell=0}^{L-1}\alphal \aab(\tbell) \dd\trp(\taul) + \NN \in \complexset{\Nrx}{SB} ~,
\end{align}
we estimate the \ac{LoS} delay and angles via MF as 
\begin{subequations} \label{eq_hat_mf}
\begin{align} \label{eq_tauhat_mf}
    \tauhatlos &= \arg \max_{\tau} \norm{ \YY \dd\conj(\tau)}_2^2 ~, \\ \label{eq_tbhat_mf}
    (\thetahazlos, \thetahellos) &= \arg \max_{\tb} \abs{\aab\herm(\tb) \YY \dd\conj(\tauhatlos)}^2 ~,
\end{align}
\end{subequations}
where searching over $\tb$ involves searching over $\thetaaz$ and $\thetael$ using \eqref{eq_ttb} (also valid hereafter\footnote{In view of \eqref{eq_ttb}, $\tb$ and $[\thetaaz, \thetael]$ will be used interchangeably throughout the text.}). The major drawback of \eqref{eq_hat_mf} is that it suffers from poor resolution in range and angle due to small bandwidth and arrays available in sub-6 GHz V2X, leading to significant estimation biases for the \ac{LoS} path in rich multipath environments. 

\subsubsection{Tensor Decomposition (CPD) Benchmark}
To surpass the resolution limitations of the MF approach, we propose to employ tensor decomposition (particularly, CANDECOMP/PARAFAC (CP) decomposition) based channel estimation to resolve \acp{MPC} \cite{tensor_overview_2017,tensor_MIMO_OFDM_2017}. With an estimated\footnote{The tensor rank is estimated using the minimum description length (MDL) criterion \cite{HOSVD_Tensor_2017}.} $L$, the CP decomposition (CPD) of $\yym$ in \eqref{eq_yy_3d} can be obtained by solving the following low-rank tensor approximation problem \cite{tensor_overview_2017}:
\begin{subequations} \label{eq_cp}
\begin{align}
    \min_{ \{\aabzl, \aabxl, \ddl \}_{\ell=0}^{L-1} } & ~\normsmall{\widehat{\yym} - \yym}_F^2
    \\
    \mathrm{s.t.}&~~  \widehat{\yym} = \sum_{\ell=0}^{L-1} \aabzl \circ \aabxl \circ \ddl ~.
\end{align}
\end{subequations} 
For ease of exposition, we introduce two \textit{multi-band selection matrices} $\JJ_{1,Q}^{B,S}=\Imatrix_B \otimes \JJ_{1,Q}^S \in \realset{(S-Q)B}{SB}$ and $\JJ_{2,Q}^{B,S}=\Imatrix_B \otimes \JJ_{2,Q}^S \in \realset{(S-Q)B}{SB}$, where
$\JJ_{1,Q}^S = \left[ \Imatrix_{S-Q} ,~ \boldzero_{(S-Q)\times Q} \right] \in \realset{(S-Q)}{S}$ and $ \JJ_{2,Q}^S = \left[ \boldzero_{(S-Q)\times Q}, ~ \Imatrix_{S-Q}   \right] \in \realset{(S-Q)}{S}$,
in which $\Imatrix_N$ is an identity matrix of size $N \times N$. Notice that $\JJ_{1,Q}^{B,S}$ and $\JJ_{2,Q}^{B,S}$ select a multi-band subarray by extracting the first and the last $S-Q$ elements, respectively, out of $S$ elements in each of the $B$ bands, resulting in a \textit{multi-band displacement invariance} structure \cite{multESP}.
Using the output vectors of the CPD in \eqref{eq_cp} and the definitions of the selection matrices, the path parameters can then be estimated  as 
$ \thetahell = \sin^{-1} ( \angle ( [\JJ_{1,1}^{1,\Mz} \aabzl]\herm [\JJ_{2,1}^{1,\Mz} \aabzl]  )/(2\pi \dz/\lambda)  )$, 
$ \thetahazl = \sin^{-1} ( {\angle ( [\JJ_{1,1}^{1,\Mx} \aabxl]\herm [\JJ_{2,1}^{1,\Mx} \aabxl]  )}/(2\pi \dx \cos(\thetahell)/\lambda) ) $ and $ \tauhatl =  -{
    \angle ( [\JJ_{1,1}^{B,S} \ddl]\herm [\JJ_{2,1}^{B,S} \ddl] )    }/( 2 \pi \deltaf) $ \cite{tensor_mMIMO_TSP_2018}. 
The \ac{LoS} parameters are obtained using\footnote{Before selecting the closest path as the LOS, a geometric consistency check based on the known \ac{CRU} height is performed to rule out non-plausible paths.}
\begin{align}\label{eq_los_cpd_3d}
    (\tauhatlos, \thetahazlos, \thetahellos) &= (\tauhat_{\ellb}, \widehat{\theta}_{\rmaz,\ellb}, \widehat{\theta}_{\rmel,\ellb}) ~,
\end{align}
where $\ellb = \arg \min_{\ell} \tauhatl$. 

\subsubsection{Tensor Decomposition with Spatial Augmentation (CPD-SA)}\label{sec_cpd_sa}
The core technical challenge with \eqref{eq_cp} pertains to a small number of antenna elements on \acp{RSU} and \acp{UE}. Typically, $\Mx \leq 4$ and $\Mz \leq 2$ in practical \ac{V2X} scenarios \cite[Table~6.1.4-5]{tr:37885-3gpp19}, while the number of \acp{MPC} can be much higher, i.e., $L \gg \max\{\Mx, \Mz\}$, especially in urban \ac{V2X} sidelink environments. In this case, the \textit{Kruskal rank condition} on the uniqueness of the CP decomposition in \eqref{eq_cp} does not hold  \cite{tensor_overview_2017,tensor_MIMO_OFDM_2017,tensor_MIMO_OFDM_2022}, leading to inaccurate estimates of the steering vectors associated with individual \acp{MPC}. Specifically, the Kruskal's condition requires that $\min\{\Mx, L\} + \min\{\Mz, L\} \geq L+2$ and $SB \geq L$ \cite{tensor_MIMO_OFDM_2022}, which is not satisfied when $L \gg \max\{\Mx, \Mz\}$. Hence, it is difficult to resolve $L$ paths with $\Mx$ or $\Mz$ antennas using a conventional CPD approach in \eqref{eq_cp}.
To circumvent the Kruskal rank problem, we propose a novel \textit{spatial augmentation} (SA) strategy for multi-band tensor-based channel estimation tailored to the unique characteristics of \ac{V2X} sidelink settings. Inspired by \cite{tensor_MIMO_OFDM_2022}, we construct a new multi-band, multi-antenna 3-D tensor $\yymt \in \complexsett{\Mz(n_z+1)}{\Mx(n_x+1)}{VB} $ from the original tensor observation in \eqref{eq_yy_3d} by augmenting the spatial domain with the frequency domain measurements, where $V \triangleq S-n_z-n_x$. The idea is to virtually increase the number of antenna elements by exploiting the fact that $S$ is large in practice (on the order of $100$s). Specifically, we construct $\yymt$ from $\yym$ in \eqref{eq_yy_3d} as\footnote{
To construct Hankel matrices, we define an operator that maps a vector $\xx \in \complexset{N}{1}$ to its Hankel matrix $\XX \in \complexset{P}{Q}$ as follows:
\begin{align}
    \XX = \hankel_P(\xx) &\triangleq \begin{bmatrix}
        x_1 & x_2 & \cdots & x_{Q} \\
        x_2 & x_3 & \cdots & x_{Q+1} \\
        \vdots & \vdots & \ddots & \vdots \\
        x_{P} & x_{P+1} & \cdots & x_{N}
    \end{bmatrix} \in \complexset{P}{Q} ~,
\end{align}
where $P+Q = N+1$ and $x_n$ denotes the $n$-th element of $\xx$.}
\begin{align} \label{eq_yymt}
    \yymt_{i, (j-1)(n_x+1)+1:j (n_x+1), (b-1)V+1:bV} = \hankel_{(n_x+1)}(\yy_{i,j,b}) \,,
\end{align}
for $i = 1, \ldots, \Mz(1+n_z)$, $j = 1, \ldots, \Mx$ and $b = 1, \ldots, B$, where we have introduced Hankel matrices constructed from $\yy_{i,j,b} \triangleq \yym_{i_1,j,(b-1)V+i_2:bV+i_2-n_z-1}$, in which 
$i_1 = \lfloor (i-1)/(n_z+1) \rfloor + 1$ and $i_2 = \text{mod}(i-1, n_z+1) + 1$. As seen from the Hankel matrix in \eqref{eq_yymt}, the proposed SA strategy performs spatial smoothing \cite{ESPRIT_Hankel_2020} for each individual band of the multi-band frequency-domain slice of $\yym$.

It can be readily verified that $\yymt$ in \eqref{eq_yymt} can be expressed as the following 3-D tensor:
\begin{align} \label{eq_yymt_3d}
     &\widetilde{\yym} =\es \sum_{\ell=0}^{L-1}\alphal \aabzw(\tbell,\taul) \circ \aabxw(\tbell,\taul) \circ \widetilde{\dd}(\taul) + \widetilde{\nnm}
 \end{align}
 where
 \begin{subequations} \label{eq_sa_steering}
 \begin{align}
     \aabzw(\tb, \tau) &= \aabz(\tb) \otimes  [\dd(\tau)]_{1:n_z+1} \in \complexset{\Mz(n_z+1)}{1} ~, \\
     \aabxw(\tb, \tau) &= \aabx(\tb) \otimes  [\dd(\tau)]_{1:n_x+1} \in \complexset{\Mx(n_x+1)}{1} ~, \\
     \widetilde{\dd}(\tau) &= \JJ_{1,S-V}^{B,S} \dd(\tau)  \in \complexset{VB}{1}~.
 \end{align}
 \end{subequations} 
With the new tensor construction in \eqref{eq_yymt_3d}, the Kruskal's condition can be satisfied if the augmentation factors $n_x$ and $n_z$ are chosen such that $\min\{\Mx (n_x+1), L\} + \min\{\Mz (n_z+1), L\} \geq L+2$ and $(S-n_z-n_x)B \geq L$. We apply CPD to $\widetilde{\yym}$ using the same framework in \eqref{eq_cp}, i.e.,
\begin{subequations} \label{eq_cp_sa}
\begin{align}
    \min_{ \{\aabzl, \aabxl, \ddl \}_{\ell=0}^{L-1} } & ~\normsmall{\widehat{\yym} - \widetilde{\yym}}_F^2
    \\
    \mathrm{s.t.}&~~  \widehat{\yym} = \sum_{\ell=0}^{L-1} \aabzl \circ \aabxl \circ \ddl ~.
\end{align}
\end{subequations} 
Based on \eqref{eq_sa_steering}, the path parameters can be estimated from the output of \eqref{eq_cp_sa} using
\begin{subequations}\label{eq_cpd_sa_est}
\begin{align} \label{eq_tauhatl_cpd_sa}
 \tauhatl &=  -\frac{
    \angle \big( [\JJ_{1,1}^{B,V} \ddl]\herm [\JJ_{2,1}^{B,V} \ddl]  \big)    }{ 2 \pi \deltaf} ~, \\
    \thetahell &= \arg \max_{\thetael} \abs{ \aabzl\herm  \aabzw(\tb, \tauhatl)   }^2 ~,    \\
    \thetahazl &= \arg \max_{\thetaaz} \abs{ \aabxl\herm  \aabxw(\tbhatel, \tauhatl)   }^2 ~,
\end{align}
\end{subequations}
where $\tbhatel \triangleq \tb(\thetaaz, \thetahell)$ (see \eqref{eq_ttb}). From the 
channel estimates of multiple paths in \eqref{eq_cpd_sa_est}, the \ac{LoS} parameters are obtained via \eqref{eq_los_cpd_3d}. We summarize the \textbf{CPD-SA} algorithm for V2X channel estimation in Algorithm~\ref{alg_cpd_3d_sa}.

\begin{algorithm}[t]
	\caption{Tensor Decomposition with Spatial Augmentation for Multi-Band V2X 3-D Channel Estimation (\textbf{CPD-SA}).}
	\label{alg_cpd_3d_sa}
	\begin{algorithmic}[1]
	    \State \textbf{Input:} Observation $\yym$ in \eqref{eq_yy_3d}.
	    \State \textbf{Output:} Delay and angle estimates of the \ac{LoS} path parameters $\{ \tauhatlos, \thetahazlos, \thetahellos\}$.
	    \begin{enumerate}[label=(\alph*)]
            \item Construct a new tensor $\yymt$ from $\yym$ using the spatial augmentation strategy in \eqref{eq_yymt}.
	        \item Compute the CPD of $\yymt$ by solving \eqref{eq_cp}.
            \item Estimate path parameters via \eqref{eq_cpd_sa_est}.
            \item Extract \ac{LoS} parameters via \eqref{eq_los_cpd_3d}. 
              
	    \end{enumerate}
	\end{algorithmic}
	\normalsize
\end{algorithm}

\subsection{ULA: 2-D Matrix Observations}\label{sec_2d_obs}
In this case, the receiver is equipped with a ULA where the antenna elements are placed horizontally, i.e., along the local X axis. Accordingly, the observation in \eqref{eq_yy_3d} degenerates to a matrix in the frequency and horizontal spatial domains:
\begin{align} \label{eq_yy_2d_ula}
     \YY =\es \sum_{\ell=0}^{L-1}\alphal   \aabx(\tbell) \dd\trp(\taul) + \NN \in \complexset{\Mx}{SB} \,.
\end{align}
Similar to the 3-D case discussed in Sec.~\ref{sec_3d_obs}, we begin by introducing benchmark algorithms for estimating \ac{LoS} parameters\footnote{Note from \eqref{eq_ax} that one can only estimate the delay $\tau$ and spatial frequency $\cos(\thetael) \sin(\thetaaz)$ of each path using \eqref{eq_yy_2d_ula}. We estimate azimuth and elevation angles from the spatial frequency estimate, assuming a known \ac{CRU} height, as discussed in Sec.~\ref{sec:RTTProtocol}.} from \eqref{eq_yy_2d_ula}, followed by the presentation of the proposed algorithm.

\subsubsection{MF Benchmark}
The MF benchmark for \eqref{eq_yy_2d_ula} applies the same steps as in \eqref{eq_hat_mf}, with the URA steering vector $\aab(\tb)$ replaced by the ULA one $\aabx(\tb)$.


\subsubsection{SVD and ESPRIT (2D-ESPRIT) Benchmark}\label{sec_svd_esp_2d}
Similar to the use of tensor decomposition in the 3-D case, we resort to \ac{SVD} and \ac{ESPRIT} to overcome the poor resolution of the MF approach. For ease of exposition, we consider the observation 
\begin{align} \label{eq_yy_trp}
    \YY\trp = \DD \Gammab \AAx\trp + \NN\trp \in \complexset{SB}{\Mx}
\end{align}
where $\DD = [ \dd(\tau_0) ~ \cdots ~ \dd(\tau_{L-1}) ] \in \complexset{SB}{L}$, 
 $\Gammab = \es \diag{ \alpha_0, \ldots, \alpha_{L-1} } \in \complexset{L}{L} $ and $\AAx = [ \aabx(\tb_0) ~ \cdots ~ \aabx(\tb_{L-1})  ] \in \complexset{\Mx}{L}$.
It is assumed that $L \leq \Mx$ and $L \leq SB$, and $\rank{\DD} = \rank{\AAx} = L$. This implies that $\rank{\YY} = L$. Let the \ac{SVD} of $\YY\trp$ be given by $ \YY\trp = \UU \Sigmab \VV\herm $, 
where $\UU \in \complexset{SB}{SB}$, $\Sigmab \in \complexset{SB}{\Mx}$ and $\VV \in \complexset{\Mx}{\Mx}$.

We first proceed with delay estimation. From the above rank assumption, there exists a non-singular matrix $\TT \in \complexset{L}{L}$ such that\footnote{The number of paths $L$ is estimated from the singular values in $\Sigmab$ using the minimum description length (MDL) criterion \cite{mdl_kailath_85}.} $\DD = \UU_{:, 1:L} \TT^{-1}$. 
On the other hand, 
following ESPRIT \cite{SVD_ESPRIT_2008},  we  exploit the multi-band displacement invariance structure of $\DD$, i.e., $\JJ_{1,1}^{B,S} \DD = \JJ_{1,2}^{B,S} \DD \Thetab $ 
where $\Thetab = \diag{ e^{\jmath 2 \pi \deltaf \tau_0} , \ldots , e^{\jmath 2 \pi \deltaf \tau_{L-1}}}$. 
Substituting $\DD = \UU_{:, 1:L} \TT^{-1}$, 
we obtain $    \JJ_{1,1}^{B,S} \UU_{:, 1:L} = \JJ_{1,2}^{B,S} \UU_{:, 1:L} \Psib$, 
where $\Psib = \TT^{-1} \Thetab \TT \in \complexset{L}{L}$. Now,  $\Psib$ can be estimated in closed-form via \ac{LS} as $\widehat{\Psib} = (\JJ_{1,2}^{B,S} \UU_{:, 1:L})^\dagger \JJ_{1,1}^{B,S} \UU_{:, 1:L}$
where $\dagger$ denotes the Moore-Penrose pseudo-inverse \cite{SVD_ESPRIT_2008}. Let $\psi_{\ell}$ be the $\ell$-th eigenvalue of $\widehat{\Psib}$. Then, the path delays can be estimated via $\tauhatl =  \angle ( \psi_{\ell} )/( 2 \pi \deltaf)$, from which 
the \ac{LoS} path is identified as $\tauhatlos = \min_{\ell} \tauhatl$. 

Next, we proceed with spatial frequency estimation. The spatial frequency of the \ac{LoS} path (i.e., $\cos(\theta_{\rm{az},0}) \sin(\theta_{\rm{el},0})$ according to \eqref{eq_ax}) can simply be estimated using the MF approach in \eqref{eq_tbhat_mf}   by $\tbhatlos = \arg \max_{\tb} \abs{\aabx\herm(\tb) \YY \dd\conj(\tauhatlos)}^2$.

\subsubsection{SVD and ESPRIT with Spatial Augmentation (2D-ESPRIT-SA)}\label{sec_svd_esp_2d_sa}
Due to small number of antenna elements $\Mx$ and large number of paths $L$ in typical V2X scenarios, the assumptions $L \leq \Mx$ and $\rank{\DD} = \rank{\AAx} = L$ do not hold in general, leading to inaccurate SVD- and ESPRIT-based estimation from \eqref{eq_yy_trp}. To restore the rank conditions, we propose to apply the 2-D version of the SA strategy in Sec.~\ref{sec_cpd_sa}. Namely, we construct a new matrix $\YYm \in \complexset{\Mx(n_x+1)}{VB} $ from the original observation in \eqref{eq_yy_trp} using 
\begin{align}  \label{eq_yymt_2d}
    \YYm_{(j-1)(n_x+1)+1:j (n_x+1), (b-1)V+1:bV} = \hankel_{(n_x+1)}(\yy_{j,b}) \,,
\end{align}
for $j = 1, \ldots, \Mx$ and $b = 1, \ldots, B$, where $V \triangleq S-n_x$ and $\yy_{j,b} \triangleq \YY_{j,(b-1)S+1:bS}$. 
The spatially augmented matrix $\YYm$ in \eqref{eq_yymt_2d} can be written in terms of new steering vectors as
\begin{align} \label{eq_yymt_2d_v2}
     &\YYm =\es \sum_{\ell=0}^{L-1}\alphal  \aabxw(\tbell,\taul)  \widetilde{\dd}\trp(\taul) + \NNm
 \end{align}
 where
 \begin{align}   
     \aabxw(\tb, \tau) &= \aabx(\tb) \otimes  [\dd(\tau)]_{1:n_x+1} \in \complexset{\Mx(n_x+1)}{1} ~, \\ \label{eq_dd_tau_new_sa}
     \widetilde{\dd}(\tau) &=  \JJ_{1,S-V}^{B,S} \dd(\tau)  \in \complexset{VB}{1} ~. 
 \end{align}
By selecting an appropriate value for $n_x$ such that both $L \leq \Mx(n_x+1)$ and $L \leq (S-n_x)B$ are satisfied, we can ensure the absence of rank-deficiency issues when estimating the parameters of the $L$ paths from $\YYm$.

We again proceed with delay estimation. We perform the same SVD-based procedure as in Sec.~\ref{sec_svd_esp_2d}, now on $\YYm$ in \eqref{eq_yymt_2d_v2}. Let $\YYm\trp = \UUm \Sigmabm \VVm\herm $
be the SVD of $\YYm\trp$, where $\UUm \in \complexset{VB}{VB}$, $\Sigmabm \in \complexset{VB}{\Mx(n_x+1)}$ and $\VVm \in \complexset{\Mx(n_x+1)}{\Mx(n_x+1)}$. Considering the structure in \eqref{eq_dd_tau_new_sa}, we estimate path delays using $\tauhatl =  \angle ( \widetilde{\psi}_{\ell} ) /( 2 \pi \deltaf)$, 
where $\widetilde{\psi}_{\ell}$ denotes the $\ell$-th eigenvalue of $(\JJ_{1,2}^{B,V} \UUm_{:, 1:L})^\dagger \JJ_{1,1}^{B,V} \UUm_{:, 1:L}$.
Following similar arguments to those in delay estimation, one can estimate the spatial frequencies $\cos(\thetahell) \sin(\thetahazl)$ using $\widehat{\omega}_{\ell} = \lambda \angle (\varphi_{\ell}) /(2\pi \dx)$, 
where $\varphi_{\ell}$ is the $\ell$-th eigenvalue of $   (\JJ_{1,2}^{1,\bar{N}_x} \VVm\conj_{:, 1:L})^\dagger \JJ_{1,1}^{1,\bar{N}_x} \VVm\conj_{:, 1:L}$, 
with $\bar{N}_x \triangleq \Mx(n_x+1)$.

Since there is no pairing between the delays $\tauhatl$ and the spatial frequencies $\widehat{\omega}_{\ell}$, we associate the \ac{LoS} delay $\tauhatlos = \min_{\ell} \tauhatl$ to a spatial frequency using the following proximity criterion: $ \whatlos = \widehat{\omega}_{\widehat{\ell}},, ~\widehat{\ell} = \arg \min_{\ell} \abs{ \widehat{\omega}_{\ell} - \whatmf }$
where $\whatmf = \arg \max_{\tb} \abs{\aabx\herm(\tb) \YY \dd\conj(\tauhatlos)}^2$ is the MF-based estimate. The proposed ESPRIT-based algorithm (\textbf{2D-ESPRIT-SA}) for 2-D channel estimation is provided in Algorithm~\ref{alg_esprit_2d_sa}.

\begin{algorithm}[t]
	\caption{ESPRIT with Spatial Augmentation for Multi-Band V2X 2-D Channel Estimation (\textbf{2D-ESPRIT-SA}).}
	\label{alg_esprit_2d_sa}
	\begin{algorithmic}[1]
	    \State \textbf{Input:} Observation  $\YY$ in \eqref{eq_yy_2d_ula}.
	    \State \textbf{Output:} Delay and spatial frequency estimates of the \ac{LoS} path parameters $\{ \tauhatlos, \whatlos \}$.
	    \begin{enumerate}[label=(\alph*)]
            \item Construct a new matrix $\YYm$ from $\YY$ using the spatial augmentation strategy in \eqref{eq_yymt_2d}. 
	        \item Compute the SVD of $\YYm\trp$.
            \item Estimate the path delays and spatial frequencies.
            \item Associate the \ac{LoS} delay estimate $\tauhatlos$ to 
            its corresponding spatial frequency estimate $\whatlos$.
	    \end{enumerate}
	\end{algorithmic}
	\normalsize
\end{algorithm}




\subsection{Single-Antenna: 1-D Vector Observations}\label{sec_1d_obs}
In the case of a single-antenna receiver, the observation in \eqref{eq_yy_3d} becomes a 1-D frequency-domain vector, given by
\begin{align} \label{eq_yy_1d}
     \yy = \es \sum_{\ell=0}^{L-1}\alphal  \dd(\taul) + \nn \in \complexset{SB}{1} \,.
\end{align}
To estimate path delays from \eqref{eq_yy_1d}, we consider two approaches, with the first one serving as benchmark.

\subsubsection{MF Benchmark}
The MF benchmark for \eqref{eq_yy_1d} simply applies $\tauhatlos = \arg \max_{\tau} \abs{ \dd\herm(\tau) \yy}^2 ~$ 
to obtain the \ac{LoS} delay.

\subsubsection{SVD and ESPRIT with Frequency Smoothing} \label{sec:1DESPRIT}
We exploit the multi-band shift invariance structure of $\dd(\tau)$ to estimate path delays from \eqref{eq_yy_1d}. Note that $\yy$ in \eqref{eq_yy_1d} can be expressed as $ \yy = \DD \alphab + \nn$, 
where $\alphab = \es [ \alpha_0, \ldots, \alpha_{L-1} ]\trp \in \complexset{L}{1}$ and $\DD$ is introduced in \eqref{eq_yy_trp}. To make the multi-band structure more evident, $\DD$ can be rewritten 
as
\begin{align} \label{eq_DD_mat2}
    \DD &= [ \dd_B(\tau_0) \otimes  \dd_S(\tau_0)  ~ \cdots ~ \dd_B(\tau_{L-1}) \otimes \dd_S(\tau_{L-1}) ] \in \complexset{SB}{L} ~
\end{align}
where where $\ddd(\tau) \in \complexset{S}{1}$ is the steering vector for a single band, i.e., $[\ddd(\tau)]_{s}=\exp(-\jmath2\pi s\Delta_{f}\tau)$, and $\dd_B(\tau) \in \complexset{S}{1}$ encodes phase shifts across consecutive frequency bands, i.e., $[\dd_B(\tau)]_{b}=\exp(-\jmath 2 \pi (f_{b-1}-f_0)\tau)$. 
To increase the rank of the single-snapshot observation in \eqref{eq_yy_1d}, we perform frequency smoothing by constructing a Hankel matrix $\YY \in \complexset{PB}{Q}$ from the observations in each individual band \cite{delayTarik_TWC_2022} $ \YY_{(b-1)P+1:bP, :} = \hankel_{P}(\yy_{(b-1)S+1:bS})$
for $b = 1, \ldots, B$, where $Q = S+1-P$. 
The smoothed matrix can be expressed as
\begin{align}\label{eq_yymt_1d_eq}
    \YY = \DD_P \Thetab_Q + \NN ~,
\end{align}
where $ \Thetab_Q \triangleq [\alphab ~ \Thetab \alphab ~ \cdots ~\Thetab^{Q-1} \alphab ] \in \complexset{L}{Q}$ ($\Thetab$ was defined after \eqref{eq_yy_trp}) and 
$ \DD_P \triangleq \Big[ \dd_B(\tau_0) \otimes  [\dd_S(\tau_0)]_{1:P}  \cdots ~ \dd_B(\tau_{L-1}) \otimes [\dd_S(\tau_{L-1})]_{1:P} \Big] \in \complexset{PB}{L}~$. 
%
Similar to the reasoning in Sec.~\ref{sec_svd_esp_2d_sa}, we can choose the stacking parameter $P$ such that $L \leq Q$ and $L \leq PB$ considering that $S \gg L$ in practice. In this case, $\rank{\Thetab_Q} = L$ and, due to the Vandermonde structure of $\dd_S(\tau)$, $\rank{\DD_P} = L$ (unless there exist two paths with exactly the same delay). The rest of the algorithm follows exactly the same procedure in Sec.~\ref{sec_svd_esp_2d}, with $\YY\trp$ in \eqref{eq_yy_trp} replaced by $\YY$ in  \eqref{eq_yymt_1d_eq} and the dimensions of all matrices adjusted accordingly (i.e., $S$ and $\Mx$ replaced by $P$ and $Q$, respectively). 

\subsection{Summary of Channel Estimation Algorithms and RTT Protocol}
In Secs.~\ref{sec_3d_obs}--\ref{sec_1d_obs}, we presented channel estimation algorithms for 3-D, 2-D and 1-D observations, respectively. These algorithms are executed at both the \ac{CRU} and \ac{RSU}, from which the \ac{RTT} and the \ac{AoA} at the \ac{RSU} are determined. These are then 
provided to the positioning methods \eqref{eq_ml}. 

To unify the nomenclature for algorithms handling observations of varying dimensions, we hereby designate the {CPD} and {2D-ESPRIT} algorithms as {HRP} (high-resolution processing). 

\section{Numerical Results}

\subsection{Scenarios}
We follow the 3GPP \ac{RSU} deployment procedure according to \cite{tr:37885-3gpp19}. Two scenarios are defined as follows. \emph{Scenario 1}: An urban scenario at an intersection with a bicycle and a car. The bicycle is driving on the bicycle lane aiming to cross the intersection. Simultaneously, another vehicle also intends to cross the same intersection. However, due to obstructing buildings, the direct visibility between the vehicle and the bicycle is either blocked or significantly hindered for a substantial duration. Therefore, a warning should be sent to the vehicle or an emergency brake should be activated in order to prevent a collision at the intersection. \emph{Scenario 2}: A  highway scenario with two cars, where one vehicle is driving straight in one lane on the highway, while another vehicle is driving on the ramp and trying to merge into the highway lane. Therefore, the first vehicle that is already on the highway, should leave sufficient space and negotiate the maneuver. In both cases, \acp{CRU} are communicating with a \ac{RSU}. 

%


\subsection{Simulation Environment}
The REMCOM Wireless InSite\textregistered  ray-tracer \cite{WirelessInSite} is utilized to simulate both scenarios. In \emph{Scenario 1}, a single \ac{RSU}, a vehicle, a bicycle, and 4 buildings are considered, as illustrated in Fig.~\ref{fig:scenarios-city}. The center of the intersection hosts the \ac{RSU}, positioned at coordinate $[0\, \text{m}, 0\, \text{m}, 10\, \text{m}]^{\mathsf{T}}$. The vehicle traverses the vehicle lane with a constant speed of $14\, \text{m/s}$ from $[1.6\, \text{m}, -70\, \text{m}, 0\, \text{m}]^{\mathsf{T}}$ to $[1.6\, \text{m}, 70\, \text{m}, 0\, \text{m}]^{\mathsf{T}}$, and its antenna positioned at a height of $1.5\, \text{m}$. On the other hand, the bicycle moves along the bicycle lane with a constant speed of $4.0\, \text{m/s}$ from $[-70\, \text{m}, -7\, \text{m}, 0\, \text{m}]^{\mathsf{T}}$ to $[70\, \text{m}, -7\, \text{m}, 0\, \text{m}]^{\mathsf{T}}$, and its antenna is located at a height of $1\, \text{m}$. Four buildings are all $50\, \text{m}$ in length, $50\, \text{m}$ in width, and $30\, \text{m}$ in height. The centers of these buildings are situated at coordinates $[\pm 45\, \text{m}, \pm 45\, \text{m}, 15\, \text{m}]^{\mathsf{T}}$, respectively. The ground is concrete and all buildings have brick walls. 
 In \emph{Scenario 2}, a single \ac{RSU}, and 2 vehicles are considered, as illustrated in Fig.~\ref{fig:scenarios-hwy}. There are 3 concrete lanes in each direction on the highway, and one lane for the entrance ramp, with each lane being $4\, \text{m}$ in width. In addition, metal traffic barriers with $0.75\, \text{m}$ in height are built on the roadside,  and grass covers the rest of the ground in the scenario. The \ac{RSU} is located in the middle of the highway with coordinate as $[0 \, \text{m},0 \, \text{m},5\, \text{m}]^{\mathsf{T}}$. The first vehicle starts at $[-360\, \text{m}, -6\, \text{m}, 0\, \text{m}]^{\mathsf{T}}$, and drives on the main lane with a constant speed of $36\, \text{m/s}$ along the x-axis to $[0\, \text{m}, -6\, \text{m}, 0\, \text{m}]^{\mathsf{T}}$. The second vehicle traverses the acceleration lane with a constant speed of $19.4\, \text{m/s}$, and it starts at $[-132\, \text{m}, -98\, \text{m}, 0\, \text{m}]^{\mathsf{T}}$, passes via $[-103\, \text{m}, -69.4\, \text{m}, 0\, \text{m}]^{\mathsf{T}}$, and then drives towards to $[0\, \text{m}, -10\, \text{m}, 0\, \text{m}]^{\mathsf{T}}$ to merge into the highway.  The antennas of both vehicles are deployed at a height of $1.5\, \text{m}$ on the top of the vehicles. 

In both scenarios, the \ac{RSU} can be deployed with a $2\times4$, a $1\times4$, or a $1\times2$ antenna array, and each \ac{CRU} can be deployed with a single antenna. 
Regarding signal parameters, the carrier frequency is $5.9~\text{GHz}$. Our approach involves using OFDM pilot signals with $12$ symbols, each having a constant amplitude. The total bandwidth is $20~\text{MHz}$, distributed over $167$ subcarriers with a subcarrier spacing of $120~\text{kHz}$, resulting in a pilot duration of $100.2~\text{us}$. The transmit power is set to $10~\text{dBm}$, while the noise spectral density is $-174~\text{dBm/Hz}$. Additionally, the receiver noise figure is calibrated at $8~\text{dB}$. RTT measurements are generated every 100 ms. In both scenarios, communication takes place between the \acp{CRU} and the \ac{RSU}, allowing the \ac{RSU} to transmit signals to the \acp{CRU} via the propagation environment in each transmission interval. However, it's essential to note that there is no direct communication established between two \acp{CRU}.
We will study different array configurations, different bandwidths, contiguity of the band, antenna patterns, and knowledge of the orientation of the \ac{RSU}. Unless otherwise specified, the default parameters are: a $2 \times 4$ array at the \ac{RSU}, a $1\times 1$ array at the \ac{CRU}, a single band with 20 MHz of bandwidth, and known \ac{RSU} orientation. 

We comprehensively assess the channel estimation results and positioning performance for each setup using the \ac{RMSE} across 100 Monte Carlo simulations. For each simulation, we firstly generate propagation paths using the raytracing data, then execute the channel estimation algorithms on the generated signals to acquire the channel parameters of \ac{LoS} paths, and finally apply the positioning algorithm 
to estimate the position of the \ac{CRU}. Here, \textbf{HRP-SA} represents the \textit{proposed} algorithm for multi-band V2X channel estimation, while \textbf{MF} and \textbf{HRP} are provided as \textit{benchmarks}. To gauge the effectiveness of our proposed bound, we also compare the \acp{RMSE} with three \acp{PEB}.  
\subsection{Channel Parameter Estimation Results}

Due to space limitations, not all combinations and  scenarios will be presented. 


\subsubsection{Performance Under Default Parameters} \label{sec_ch_perf_default} 
\begin{figure}
     \centering
     \subfigure[\ac{CDF} of absolute range error.]{         \input{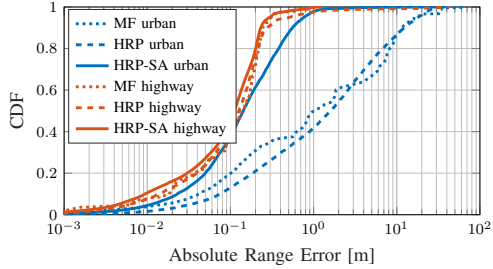}
         \label{fig:figure1_c_e}}
     \\
     \subfigure[\ac{CDF} of absolute azimuth error.]{
     \input{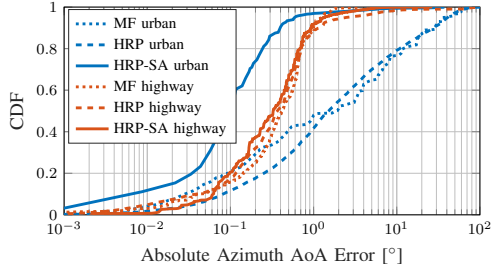}
         \label{fig:figure2_c_e}
     }
     \\
     \subfigure[\ac{CDF} of absolute elevation error.]{
     \input{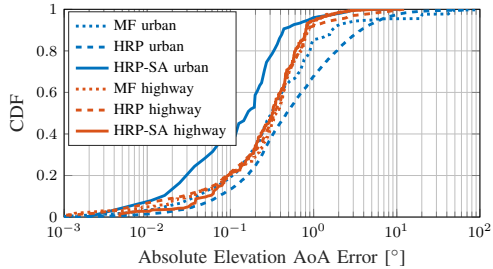}
         \label{fig:figure3_c_e}
     }
     \caption{Comparative channel estimation performance of the considered algorithms  for the urban vehicle scenario (urban) and the merging vehicle in the highway scenario (highway) with default parameters.}
     \label{fig:all_figures_channel_estimation}
 \vspace{-5mm} \end{figure}

We run the three channel estimation algorithms 
on the received signals to obtain the \ac{LoS} parameter estimates. 
Fig.~\ref{fig:all_figures_channel_estimation} illustrates the \ac{CDF} performance of the considered algorithms in Scenario~1 (urban, vehicle) and Scenario~2 (highway, merging vehicle). These results are aggregated over both the \ac{CRU} (respectively \ac{CRU}) trajectory and multiple Monte Carlo noise realizations, and presented in terms of absolute errors for range, azimuth, and elevation. From Fig.~\ref{fig:all_figures_channel_estimation}, we observe that in the urban scenario the proposed HRP-SA algorithm provides significant performance gains over the benchmarks MF and HRP (approximately two and one order-of-magnitude accuracy improvement in range/azimuth and elevation, respectively, considering $90\%$ percentile errors). This demonstrates the effectiveness of the proposed SA strategy in resolving the \ac{LoS} path from the \ac{NLoS} paths, particularly in the context of rich multipath environments commonly encountered in V2X sub-6 GHz propagation scenarios (such as in the urban scenario). Conventional MF and HRP approaches fail to distinguish closely spaced paths with small bandwidths and the low number of antenna elements available in V2X sub-6 GHz. On the other hand, in the highway scenario, all the algorithms exhibit similar performance. This scenario involves a much sparser multipath profile, primarily \ac{LoS}-dominant propagation with few weak \ac{NLoS} paths. As a result, the MF approach, which essentially discards the existence of \ac{NLoS} paths, performs equally well as the HRP-SA algorithm.





\subsubsection{Illustrative Example on Comparative Performance of Algorithms} 
\begin{figure}
     \centering
\input{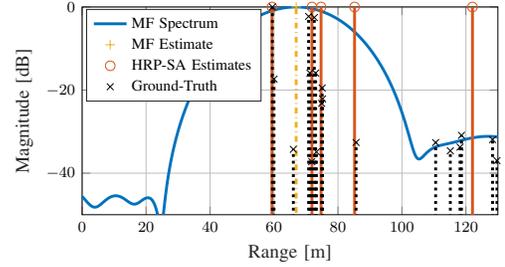}
     \caption{The range profile obtained by the MF approach for the CRU located at $-58.8$ m in the urban vehicle scenario, along with the range estimates of HRP-SA and the ground-truth ranges of multiple paths.}
     \label{fig:all_figures_illustrative}
 \vspace{-5mm} \end{figure}

To demonstrate how HRP-SA outperforms MF in the urban scenario, we provide an illustrative example in Fig.~\ref{fig:all_figures_illustrative}, where the range profile in \eqref{eq_tauhat_mf} obtained by the MF approach is plotted together with the HRP-SA range estimates and the ground-truth path ranges. It is observed that the proposed high-resolution channel estimator can pinpoint the location of the \ac{LoS} component in range, while the conventional MF approach leads to a large bias due to the merging of the \ac{LoS} path and the closely spaced \ac{NLoS} paths in low-bandwidth sub-6 GHz operation. This example demonstrates the significance and benefits of employing high-resolution processing in dense multipath environments, typically in urban V2X sidelink scenarios.

\subsubsection{Impact of RSU Array Configurations}
\begin{figure}
     \centering
     \subfigure[\ac{CDF} of absolute range error.]{
         \input{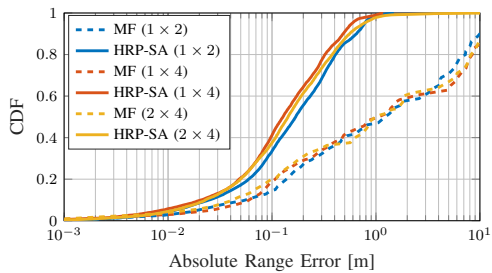}
         \label{fig:figure1_a_p}
     }\vspace{-1mm} 
     \\
     \subfigure[\ac{CDF} of absolute azimuth error.]{
         \input{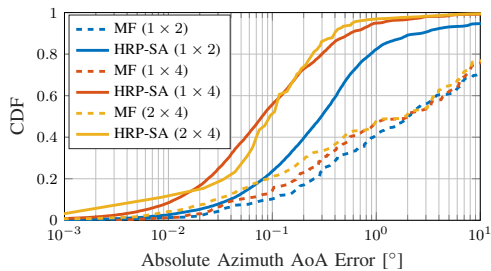}
         \label{fig:figure2_a_p}
     }
     \vspace{-1mm} 
     \caption{Channel estimation performance of MF and HRP-SA algorithms for various RSU array configurations for the urban vehicle scenario. 
     }
     \label{fig:all_figures_antenna_pattern}
 \vspace{-5mm} \end{figure}

Next, we analyze the effect of antenna array configurations on channel estimation 
and compare three different array configurations at the RSU side, $1\times2$, $1\times4$, and $2\times4$. The results are summarized in Fig.~\ref{fig:all_figures_antenna_pattern}. We note that the algorithms exploit the prior information on the CRU height for the $1\times2$ and $1\times4$ ULA configurations to obtain azimuth and elevation angles from spatial frequency estimates, while the $2\times4$ URA relies only on observations to estimate angles. Fig.~\ref{fig:all_figures_antenna_pattern} illustrates that, as a general trend, employing a larger antenna array size leads to improved channel estimates for the proposed HRP-SA estimator.\footnote{As a noteworthy exception, it is observed that HRP-SA with a $2\times4$ array yields a comparatively worse elevation performance compared to $1\times2$ and $1\times4$ arrays. This anomaly can be attributed to the potential superior informativeness of prior information regarding the CRU height when compared to the observations from the URA.} This results from the enhanced angular resolution achieved with larger arrays, which, in turn, facilitates improved path resolvability in both the range and angle domains. Hence, HRP-SA effectively exploits the available resolution capabilities in both domains to separate the \ac{LoS} path from closely spaced \ac{NLoS} paths. 
On the other hand, the MF algorithm exhibits similar performance across the different array configurations, with the gains associated with larger arrays not as prominently observed as in the case of HRP-SA. The reason is that MF treats all three dimensions separately and performs sequential processing to extract \ac{LoS} parameters (i.e., first, delay estimation via \eqref{eq_tauhat_mf} and then angle estimation via \eqref{eq_tbhat_mf}). This limitation of the MF approach impedes full utilization of available resolution, resulting in smaller gains in channel estimation as array size increases.

\subsubsection{Impact of Bandwidth and Number of Bands} \label{multiband_channel_estimation}
\begin{figure}[t]
    \centering
     \subfigure[Single-band performance on range estimation for 20 (blue), 40 (red), 60 (orange), and 80 (purple) MHz.]{
\input{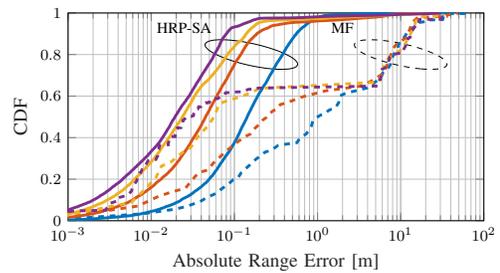}
    \label{fig:cdf_compare_bw}
    }
    \\
\subfigure[Single-band performance on AoA azimuth estimation for 20 (blue), 40 (red), 60 (orange), and 80 (purple) MHz.]
{\input{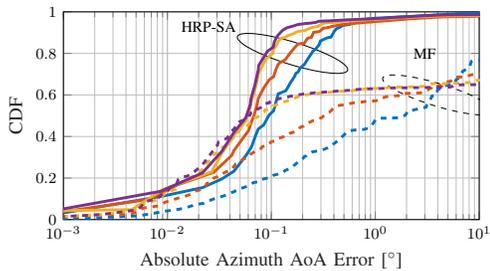}
    \label{fig:cdf_compare_bw_az}
    }
    \\
\subfigure[Multi-band performance on range estimation for $B=1$ (blue), $B=2$ (red), $B=3$ (orange), and $B=4$ (purple).]{   
         \input{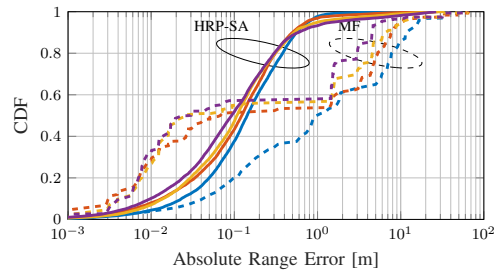}
    \label{fig:cdf_nbands}
    }
\\
\subfigure[Multi-band performance on AoA azimuth estimation for $B=1$ (blue), $B=2$ (red), $B=3$ (orange), and $B=4$ (purple).]{    
         \input{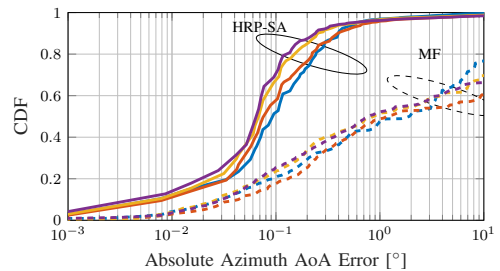}
         \label{fig:cdf_nbands_az}
}
      \caption{Analysis of the bandwidth and the number of bands $B$ on range estimation performance with a fixed frequency separation of $100~\rm{MHz}$ between the consecutive bands for the urban vehicle scenario.}
    \label{fig:totalBandwidth}
\vspace{-5mm} \end{figure}

We evaluate the impact of bandwidth on the performance of range estimation in a single-band scenario by using four different values of single-band bandwidths, i.e., $20~\rm{MHz}$, $40~\rm{MHz}$, $60~\rm{MHz}$ and $80~\rm{MHz}$. The results are shown in Fig.~\ref{fig:cdf_compare_bw}. We observe that both the benchmark MF approach and the proposed HRP-SA algorithm provide more accurate \ac{LoS} range estimates as the bandwidth increases.\footnote{It is crucial to emphasize that the total power remains constant while the bandwidth changes. Therefore, any enhancements in accuracy are solely attributed to the increase in bandwidth.} This is expected as a larger bandwidth inherently translates to superior range resolution, thus facilitating improved path resolvability. Similar improvements can be observed for azimuth estimation in Fig.~\ref{fig:cdf_compare_bw_az}.\footnote{By examining both Fig.~\ref{fig:figure2_a_p} and Fig.~\ref{fig:cdf_compare_bw_az} together, we can infer that the MF approach benefits more from increasing bandwidth in terms of azimuth estimation accuracy compared to increasing array size. This observation results from the fact that the MF approach initially processes the delay dimension.}
%
%
%
In Fig.~\ref{fig:cdf_nbands}, we assess the range estimation performance of MF and HRP-SA in multi-band operation by varying the number of bands $B$ from $1$ to $4$ with a fixed frequency separation of $100~\rm{MHz}$ between the consecutive bands.\footnote{As in the scenario involving variable bandwidths, we maintain a constant total power while altering the number of bands. This approach enables us to isolate and analyze the specific impact of varying the number of bands, independent of any power increase.} It is seen that as a general trend, increasing the number of bands improves \ac{LoS} ranging accuracy for both algorithms since we obtain larger frequency apertures with higher number of bands, leading to enhanced range resolution \cite{delayTarik_TWC_2022}. Finally, Fig.~\ref{fig:cdf_nbands_az} shows the azimuth estimation performance in multi-band operation, which reveals similar trends as in range estimation.

\subsection{Positioning Results}


    






We proceed with the results of positioning, which utilizes the output of channel estimation routines.

\subsubsection{RMSE and CDF under Default Parameters}
\begin{figure}
    \centering
 \subfigure[Positioning RMSE of the vehicle position estimates using different channel estimation results and predicted positioning RMSE from the bounds for the urban vehicle scenario.]{
  \input{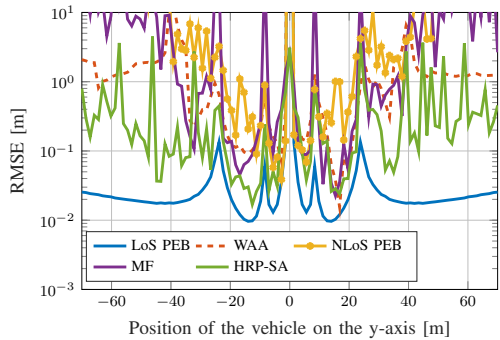}
\label{fig:urban_scenario_performance}
    }\vspace{-1mm} 
    \\
 \subfigure[Comparison between the \acp{CDF} of the absolute positioning error using MF and HRP-SA channel estimators for the vehicle and bicycle in the urban scenario.]{
  \input{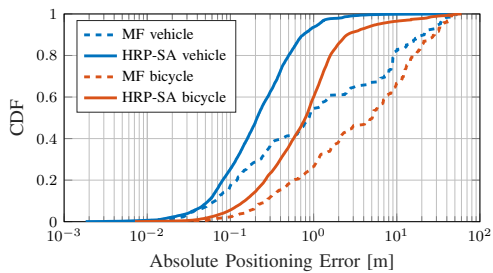}
\label{fig:all_urban_scenario}
   }\vspace{-1mm} 
   \\
    \subfigure[Comparison between the \acp{CDF} of the absolute positioning error when using MF and HRP-SA channel estimators for the vehicles in the highway scenario.]{
  \input{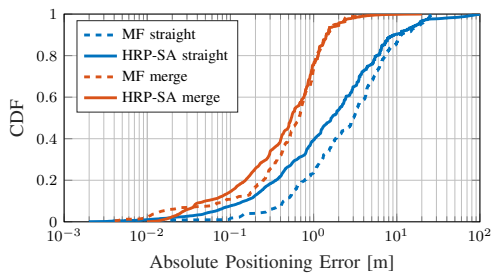}
\label{fig:all_highway_scenario}
   }\vspace{-1mm} 
   \caption{RMSE and CDF performances comparison when using MF and HRP-SA channel estimators and the default parameters for both vehicle and highway scenarios.}
\label{fig:highway_scenario_performance-combined}
\vspace{-5mm} \end{figure}

We firstly utilize the default parameters, and investigate the positioning RMSE performance  of MF and HRP-SA for the urban vehicle scenario, as shown in Fig.~\ref{fig:urban_scenario_performance}. We also compare the RMSE performance with 3 bounds from Section \ref{sec:CRB-proposed}: (i) the \ac{PEB} considering estimation of only the \ac{LoS} path parameters between \ac{CRU} and \ac{RSU} (denoted by LoS PEB), (ii) the \ac{PEB} considering estimation of all the paths in the  \ac{LoS} resolution cell, 
    and (iii) the  proposed \ac{WAA}. 
 In terms of the algorithms, we observe that HRP-SA significantly outperforms MF, as expected from the previous section. 
A moderate accuracy requirement (1–3 m) can be attained by HRP-SA throughout the trajectory and sub-meter accuracy can be achieved for most positions on the trajectory, except for a few positions where the multipath is highly complicated and the inter-path interference is serious. 
    Regarding the three bounds, the LoS-only \ac{PEB} solely considers the \ac{LoS} path, thus exhibiting an excessively optimistic outlook. Conversely, the \ac{NLoS} \ac{PEB} considers all path parameters, which leads to overly pessimistic results. Notably, the disparity between these two bounds is substantial. In contrast, the  \ac{WAA}  accounts for the challenge of non-resolvability in path determination, thus providing more reasonable bounds that effectively align with the performance of the MF estimator.  

We then utilize the same configuration, study the \acp{CDF} of the absolute positioning error for all four scenarios, and compare the positioning performance when using MF and HRP-SA channel estimation results, as indicated in Fig.~\ref{fig:all_urban_scenario} and Fig.~\ref{fig:all_highway_scenario}. From Fig.~\ref{fig:all_urban_scenario}, we observe that  HRP-SA results in better positioning performance than the MF for both vehicle and bicycle in the urban scenario.
Moreover, sub-meter accuracy can be achieved at 93.5\% for the urban vehicle scenario and moderate accuracy (error below 3 m) can be 91.4\% for the urban bicycle scenario when the HRP-SA channel estimator is implemented. 
From Fig.~\ref{fig:all_highway_scenario}, we observe that HRP-SA results in better positioning performance than  MF for both vehicles in the highway scenario. Moderate accuracy can be achieved  65\% (resp.~98.1\%) of the time for the straight-moving (resp.~merging) vehicle with HRP-SA. This is because  the straight-moving vehicle starts far away from the RSU, where small \ac{AoA} estimation errors bring  significant positioning errors. The gap between HRP-SA and MF is smaller than in the urban scenario since the multipath is far less complicated, as discussed in Sec.~\ref{sec_ch_perf_default}. 



\subsubsection{Impact of RSU Array Configurations}
\begin{figure}
    \centering
    \input{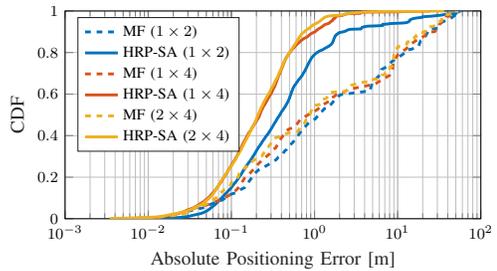}\vspace{-1.5mm} 
    \caption{Analysis of the RSU array configuration on positioning performance for the urban vehicle scenario.}
 \label{fig:compare_antenna_pattern}
\vspace{-5mm} \end{figure}

We again focus on the urban vehicle scenario,  and study the effect of the array configurations on the positioning performance, where we compare three different antenna configurations for the \ac{RSU}, $1\times2$, $1\times4$, and $2\times4$, and use all the rest of default parameters. \acp{CDF} of the absolute positioning error for three antenna configurations are summarized in Fig.~\ref{fig:compare_antenna_pattern}. We find that the larger the antenna array is used, the better positioning performance can be achieved. This is because a larger antenna array can provide higher angular resolution, resulting in more accurate angle and range estimates as demonstrated in Fig.~\ref{fig:all_figures_antenna_pattern}, which improves the positioning performance of the system, especially at long distances where the positioning performance is sensitive to the angular error. 


\subsubsection{Impact of Bandwidth and Number of Bands}
\begin{figure}
    \centering
    \subfigure[Single band with different bandwidths $S \Delta_f$ for 20 (blue), 40 (red), 60 (orange), and 80 (purple) MHz.]{
     \input{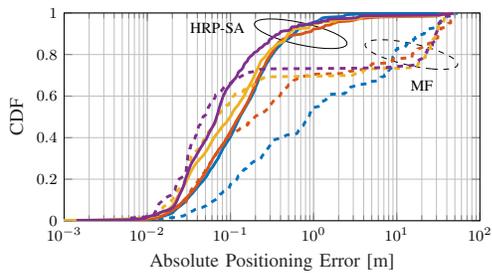}
\label{fig:compare_bandwidth}
    }\vspace{-1mm} 
\\    
    \subfigure[Fixed band separation of 100 MHz with different number of bands $B$ for $1$ (blue), $2$ (red), $3$ (orange), and $4$ (purple).]{
 \input{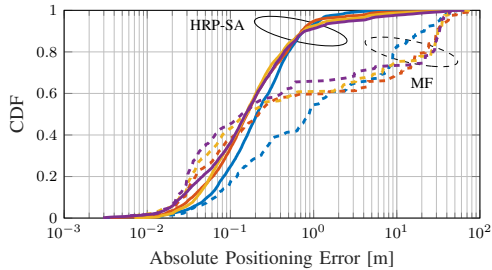}
\label{fig:compare_number_of_bands}
    }\vspace{-1mm} 
    \\
    \subfigure[Two bands with a fixed single bandwidth, and different band separations for 20 (blue), 100 (red), 200 (orange), and 400 (purple) MHz.]{
  \input{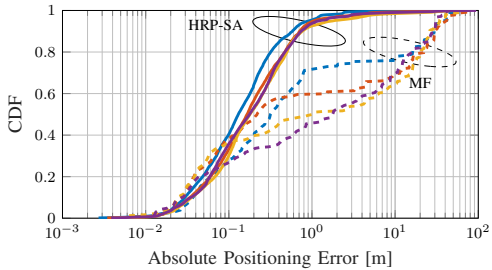}
\label{fig:frequency_separation}
    }
    \vspace{-1.5mm} 
    \caption{Analysis of the bandwidth, the number of bands $B$, and the band separation between the consecutive bands on positioning performance for the urban vehicle scenario.}
    \label{fig:enter-label}
\vspace{-2mm} \end{figure}

Next, the effect of the size of bandwidth on the positioning performance is investigated, where four different values of single-band  bandwidths, 20 MHz, 40 MHz, 60 MHz, and 80 MHz are used in the respective order, and the results are displayed in Fig.~\ref{fig:compare_bandwidth}. 
Overall, we observe that larger bandwidths lead to higher positioning accuracy. The reason is that increased bandwidth translates to higher delay resolution, resulting in more accurate range and angle estimates, as demonstrated in Fig.~\ref{fig:cdf_compare_bw} and Fig.~\ref{fig:cdf_compare_bw_az}. Additionally, we note that for small positioning errors at the centimeter level, the MF approach can outperform HRP-SA. This occurs in instances characterized by \ac{LoS}-dominant propagation in the trajectory, where the MF benchmark exhibits asymptotic optimality at high SNRs, achieving the \ac{LoS} PEB, as indicated by the results around $0 \, \rm{m}$ in Fig.~\ref{fig:urban_scenario_performance}. However, for larger errors at the decimeter and meter levels, typically occurring under dense multipath conditions (e.g., outside the interval $[-20 , 20] \, \rm{m}$ in Fig.~\ref{fig:urban_scenario_performance}), HRP-SA significantly outperforms MF, particularly when path resolvability becomes critical.
Fig.~\ref{fig:compare_number_of_bands} compares the positioning performance of the system using different numbers of bands $B$ from 1 to 4 with a  fixed frequency separation of 100 MHz between the consecutive bands. From the figure, we can summarize that a larger number of  bands in general benefits the \ac{CRU} positioning, since larger frequency apertures can be obtained through more bands, which bring better range estimates as explained in Section \ref{multiband_channel_estimation}. In addition, results for different band separations are shown in Fig.~\ref{fig:frequency_separation}, where two bands are used and the separations between two bands are 20 MHz, 100 MHz, 200 MHz, and 400 MHz, respectively. We observe that increasing the band separation does not always improve the positioning accuracy. This is because increased separation in the frequency aperture results in range ambiguities, which leads to selection of spurious peaks in range estimation and thus degradation of positioning performance, in alignment with the observations in \cite{delayTarik_TWC_2022}. As a rough conclusion, larger contiguous bands and smaller separations between the consecutive bands are preferable to achieve better positioning performance. 


\subsubsection{Impact of RSU Heading Uncertainty}
\begin{figure}
    \centering
    \input{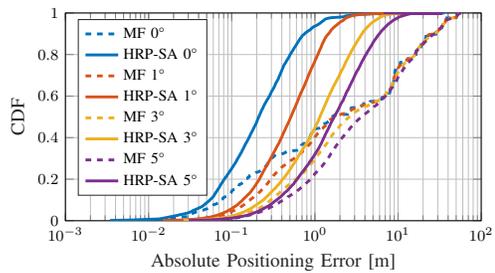}
    \vspace{-1.5mm} 
    \caption{Analysis of the RSU heading uncertainty on positioning performance for the urban vehicle scenario.}
\label{fig:angular_uncertain}
\vspace{-5mm} \end{figure}

So far, we assumed the orientation of the \ac{RSU} is perfectly known for all the above results. However, it is an ideal assumption, and the orientation is usually provided with some uncertainties. We then utilize the default parameters again, but add disturbances to the heading, where the additive disturbances follow zero-mean Gaussian distributions with standard deviations as 1\textdegree, 3\textdegree, and 5\textdegree, respectively. The corresponding positioning performances are displayed in Fig.~\ref{fig:angular_uncertain}, compared with the benchmark (the heading is perfectly known). 
From Fig.~\ref{fig:angular_uncertain}, we can clearly observe that the heading uncertainty degrades the positioning performance, and a large uncertainty causes a worse performance. The reason is obvious: the azimuth of the \ac{AoA} is affected by the heading, which is used to localize the \ac{CRU}. A small error on the \ac{AoA} can cause a significant positioning error, especially at a long distance. {Therefore, careful calibration on the RSU orientation should be performed.}

\vspace{-2mm} 
\section{Conclusions}

In this  paper, our primary focus is on the examination of sidelink V2X positioning within the context of 3GPP Release 18, specifically within the FR1 frequency range, with the aim of localizing a \ac{CRU} with the aid of a single multi-antenna \ac{RSU}. We introduce high-resolution channel parameter estimation algorithms for both URAs and ULAs to estimate delay, azimuth and elevation angles of multipath components, suitable for multiband operation. 
In addition, we also generalize the CRB on the range in \cite{ge2023analysis} to an innovative methodology that leverages Fisher information analysis to anticipate the performance of positioning in a multipath propagation environment. Furthermore, we assess the channel estimators, the sidelink positioning system itself, and the predictive performance of our methodology across an urban and a highway scenario, employing ray-tracing data.

Our findings reveal that the standard MF channel estimator is limited in resolution, as corroborated by the new approximate performance bound, thus motivating the need for super-resolution methods. 
The proposed high-resolution algorithms 
are demonstrated to work well in both URA and ULA cases, support multiband operation, and provide accurate estimations of delays, and azimuth and elevation angles, as well as the positions, in most cases outperforming the standard MF channel estimator. 
We quantify the benefit of larger antenna arrays to enhance the performance, by harnessing increased angular resolution, which in turn also improves delay estimation. The benefit of a larger bandwidth in multiband processing is less obvious: while a larger contiguous band always improves performance, phase-coherent sub-bands are mainly beneficial for improving smaller errors (while increasing larger errors) and require careful spacing to avoid degradation due to ambiguities. Finally, we revealed the importance of careful calibration by showing that even small orientation errors at the \ac{RSU} cause severe positioning degradation. 


Despite the complexities of the scenarios and the severe multipath interference, our research demonstrates that achieving sub-meter positioning accuracy is feasible, even with moderate array sizes at the \ac{RSU} (e.g., $1\times4$ or $2\times 4$) and moderate bandwidths (20--40 MHz). 


\section*{Acknowledgments}
The authors wish to thank Remcom for providing Wireless InSite\textregistered  ray-tracer.

\balance
\bibliography{IEEEabrv_paper,Bibl_paper}

\end{document}